\documentclass[prc,twocolumn,showpacs,preprintnumbers,amsmath,amssymb,superscriptaddress,floatfix]{revtex4-2}

\usepackage{graphicx}% Include figure files
\usepackage{subfigure}
\usepackage{dcolumn}% Align table columns on decimal point
\usepackage{bm}% bold math
\usepackage[dvipsnames]{xcolor}
\colorlet{RED}{red}
\usepackage[T1]{fontenc}
\usepackage[utf8]{inputenc}
\usepackage{hyperref}
\usepackage{textcomp}

\hypersetup{
colorlinks,
linkcolor={red},
citecolor={blue},
urlcolor={blue}
}

\begin{document}

%\title{Emergence of axial vs. triaxial shape coexistence in neutron-rich Nb isotopes: new insights from high-resolution spectroscopy of odd-even \texorpdfstring{$^{105-109}$}{105-109}Nb isotopes}
%\title{Emergence of axial vs. triaxial shape coexistence in neutron-rich odd-even \texorpdfstring{$^{105-109}$}{105-109}Nb isotopes}
\title{Shape evolution in neutron-rich odd-even \texorpdfstring{$^{105-109}$}{105-109}Nb isotopes}

\newcommand{\IPHC}{Universit\'e de Strasbourg, CNRS, IPHC UMR 7178, F-67000 Strasbourg, France}
\newcommand{\IPTwoI}{Universite Claude Bernard Lyon 1, CNRS/IN2P3, IP2I Lyon, UMR 5822, Villeurbanne, F-69100, France}
\newcommand{\GANIL}{GANIL, CEA/DRF-CNRS/IN2P3, BP 55027, 14076 Caen cedex 5, France}
\newcommand{\CSNSM}{CSNSM, Univ. Paris-Sud, CNRS/IN2P3, Universit\'e Paris-Saclay, 91405 Orsay, France}
\newcommand{\TUDarmstadt}{Institut f\"ur Kernphysik, Technische Universit\"at Darmstadt, D-64289 Darmstadt, Germany}
\newcommand{\IFIC}{Instituto de F\'isica Corpuscular, CSIC-Universitat de Val\`encia, E-46980 Valencia, Spain}
\newcommand{\Canada}{Department of Chemistry, Simon Fraser University, Burnaby, British Columbia, Canada}
\newcommand{\Legnaro}{INFN, Laboratori Nazionali di Legnaro, Via Romea 4, I-35020 Legnaro, Italy}
\newcommand{\IPN}{Institut de Physique Nucl\'eaire, IN2P3-CNRS, Univ. Paris Sud, Universit\'e Paris Saclay, 91406 Orsay Cedex, France}
\newcommand{\Debrecen}{Institute for Nuclear Research of the Hungarian Academy of Sciences, Pf.51, H-4001, Debrecen, Hungary}
\newcommand{\Somerset}{iThemba LABS, National Research Foundation, P.O.Box 722, Somerset West,7129 South Africa}
\newcommand{\Padova}{INFN Sezione di Padova, I-35131 Padova, Italy}
\newcommand{\UPadova}{Dipartimento di Fisica e Astronomia dell'Universit\`a di Padova, I-35131 Padova, Italy}
\newcommand{\GSI}{GSI, Helmholtzzentrum f\"ur Schwerionenforschung GmbH, D-64291 Darmstadt, Germany}
\newcommand{\Milano}{INFN, Sezione di Milano, Milano, Italy}
\newcommand{\LPSC}{LPSC, Universit\'e Grenoble-Alpes, CNRS/IN2P3, 38026 Grenoble Cedex, France}
\newcommand{\IRFU}{IRFU, CEA/DRF, Centre CEA de Saclay, F-91191 Gif-sur-Yvette Cedex, France}
\newcommand{\UMilano}{Dipartimento di Fisica, Universit\`a di Milano, I-20133 Milano, Italy}
\newcommand{\STFC}{STFC Daresbury Laboratory, Daresbury, Warrington, WA4 4AD, UK}
\newcommand{\ILL}{Institut Laue-Langevin, Grenoble, France}
\newcommand{\IFINHH}{IFIN-HH, Romania}
\newcommand{\Vanderbilt}{Department of Physics and Astronomy, Vanderbilt University, Nashville, Tennessee 37235, USA}
\newcommand{\VECC}{Variable Energy Cyclotron Centre, 1/AF Bidhannagar, Kolkata 700064, India}
\newcommand{\HBNI}{Homi Bhabha National Institute, Training School Complex, Anushaktinagar, Mumbai-400094, India}
\newcommand{\Jinan}{Shandong Provincial Key Laboratory of Nuclear Science, Nuclear Energy Technology and Comprehensive Utilization, Weihai Frontier Innovation Institute of Nuclear Technology, School of Nuclear Science, Energy and Power Engineering, Shandong University, Jinan 250061, China}
\newcommand{\Weihai}{Weihai Research Institute of Industrial Technology of Shandong University, Weihai 264209, China}
\newcommand{\Daejeon}{Center for Exotic Nuclear Studies, Institute for Basic Science, Daejeon 34126, Republic of Korea}
\newcommand{\Berkeley}{Lawrence Berkeley National Laboratory, Berkeley, California 94720, USA}
\newcommand{\Beijing}{Department of Physics, Tsinghua University, Beijing 100084, China}

%\collaboration{MUSO Collaboration}%\noaffiliation

%First authors
%M.~Abushawish, E.~H.~Wang, J.~Dudouet, A.~Navin
% Gammasphere
%J.H.Hamilton, A.V.Ramayya, Y.X.Luo, J.O.Rasmussen, S.J.Zhu.
% E680
% F. Didierjean, G. Duchene, C. Michelagnoli, M. Rejmund, A. Lemasson, E. Cl\'ement,
% E661
% S. Bhattacharyya, M. Rejmund, A. Lemasson, E. Cl\'ement, Yung Hee, Christelle schmidt, Bertrand

%First authors

\author{M.~Abushawish} %mogahedsameerq@gmail.com
\altaffiliation{Present address: Grupo de Física Nuclear , Universidad Complutense de Madrid, CEI Moncloa, 28040 Madrid, Spain}
\affiliation{\IPTwoI}

\author{E.~H.~Wang}
\email[email: ]{enWang9864@outlook.com}
\affiliation{\Jinan}
\affiliation{\Weihai}
\affiliation{\Vanderbilt}

\author{J.~Dudouet}
\email[Corresponding author: ]{j.dudouet@ip2i.in2p3.fr}
\affiliation{\IPTwoI}

\author{A.~Navin} %navin.alahari@ganil.fr
\affiliation{\GANIL}

%second author list (Spokesperson, instrument responsible)

\author{E. Clément} % clement@ganil.fr
\affiliation{\GANIL}

\author{G. Duchêne} % gilbert.duchene@iphc.cnrs.fr
\affiliation{\IPHC}

\author{J.~H.~Hamilton} % j.h.hamilton@Vanderbilt.Edu
\affiliation{\Vanderbilt}

\author{A. Lemasson} % antoine.lemasson@ganil.fr
\affiliation{\GANIL}

\author{C.~Michelagnoli} % michelagnolic@ill.fr
\altaffiliation{Present address: Institut Laue-Langevin, F-38042 Grenoble Cedex, France}
\affiliation{\GANIL}

% \author{M. Rejmund} % maurycy.rejmund@ganil.fr
% \affiliation{\GANIL}

\author{O. Stezowski} % o.stezowski@ip2i.in2p3.fr
\affiliation{\IPTwoI}%

%third author list 

\author{S. Bhattacharyya} % sarmi@vecc.gov.in
\affiliation{\VECC}
\affiliation{\HBNI}

\author{F. DidierJean} % francois.didierjean@iphc.cnrs.fr
\affiliation{\IPHC}

\author{B.~Jacquot} %bertrand.jacquot@ganil.fr 
\affiliation{\GANIL}

\author{Y. H. Kim} %yunghee.kim@ibs.re.kr
\altaffiliation{Present address: Center for Exotic Nuclear Studies, Institute for Basic Science, Daejeon 34126, Republic of Korea}
\affiliation{\GANIL}

\author{Y.~X.~Luo} %haishi306@gmail.com
\affiliation{\Berkeley}

\author{A.~V.~Ramayya} %a.v.ramayya@Vanderbilt.edu
\affiliation{\Vanderbilt}

\author{J.~O.~Rasmussen} %oxras1@gmail.com
\affiliation{\Berkeley}

% \author{C.~Schmitt} %christelle.schmitt@iphc.cnrs.fr
% \altaffiliation{Present address: IPHC Strasbourg, Universit\'e de Strasbourg-CNRS/IN2P3, F-67037 Strasbourg Cedex 2, France}
% \affiliation{\GANIL} 

\author{S.~J.~Zhu} %zhushj@mail.tsinghua.edu.cn
\affiliation{\Beijing}

\date{\today}

\begin{abstract}

\textbf{Background:} Neutron-rich nuclei around $Z\sim40$ are well known for exhibiting multiple shape transitions. This region shows one of the sharpest shape transitions in the nuclear chart, evolving from a spherical vibrator at $N=58$ to a strongly deformed prolate shape at $N=60$. The largest deformations are observed for $_{38}$Sr and $_{40}$Zr. This abrupt shape transition disappears at $Z=36$ and below, where a shape transition from spherical to oblate nuclei is predicted. On the other hand, for $Z\ge42$ and $N\ge60$, the shape is known to evolve from axial to triaxial. While the even-$Z$ nuclei in this region have already been extensively studied, new insights can be gained from the studies of odd-$Z$ isotopes for a better understanding of the underlying mechanisms driving these phenomena.\\
\textbf{Purpose:} 
The $_{41}$Nb nuclei lie at the boundary between axially deformed Zr and triaxially deformed Mo nuclei. This work investigates the nuclear structure of very neutron-rich Nb nuclei up to $N=68$. The goal is to understand how the nuclear shape evolves as a function of isospin in this isotopic chain and provide new insights into the emergence of triaxial deformation.\\
\textbf{Methods:} 
The structure of the neutron-rich Nb isotopes was investigated using state-of-the-art high-resolution $\gamma$-ray spectroscopy of fission fragments produced via two different fission reactions. The use of $^{9}$Be($^{238}$U,f) inverse kinematics, with a detection system comprising AGATA, EXOGAM, and VAMOS++, enabled the measurement of prompt and delayed $\gamma$ rays from isotopically identified fission fragments and $\gamma-\gamma-\gamma-\gamma$ high-fold data were obtained from a spontaneous fission source of $^{252}$Cf using the Gammasphere array.\\
\textbf{Results:} 
The level scheme of $^{105}$Nb has been significantly extended, with the addition of two negative-parity bands observed for the first time. A new level scheme is proposed for $^{107}$Nb, which is not in agreement with an earlier measurement, and new levels and transitions have been added to the very neutron-rich $^{109}$Nb. The degree of triaxiality of the new bands is discussed on the basis of signature splitting analysis. The recently reported level scheme of $^{99}$Nb has been revised.\\
\textbf{Conclusions:} 
This systematic study on the Nb isotopic chain, compared to Zr and Mo, indicates that while the ground-state band exhibits a triaxial deformation, attributed to a proton hole coupled to a triaxially deformed Mo core, the negative-parity bands, based on isomeric band-heads, display an axially symmetric deformed structure, similar to that observed in the Zr isotopes, revealing the existence of a shape coexistence in the neutron-rich Nb nuclei.
\end{abstract}

\maketitle

\section{Introduction}
\label{sec:introduction}

The study of neutron-rich nuclei in the region around $Z\sim40$ has revealed a complex interplay of nuclear shapes~\cite{Heyde2011,Garrett2022} and remains a major focus of both experimental and theoretical efforts. One of the most pronounced shape transitions in the nuclear chart occurs as nuclei evolve from spherical configurations at $N=58$ to strongly axially deformed shapes at $N=60$, interpreted as a quantum phase transition~\cite{Togashi2016}. This transition reaches its maximum intensity in Sr ($Z=38$) and Zr ($Z=40$) nuclei. However, this phenomenon has been shown to vanish for the low-$Z$ side of this island of deformation, as observed in Kr ($Z=36$)~\cite{Dudouet2017,Flavigny2017}, where a transition toward slightly deformed oblate shapes is predicted~\cite{Rodriguez2014,Flavigny2017,Dudouet2024}. On the high-$Z$ side, the nuclear shape evolves from axial in Zr to triaxial in Mo~\cite{Abusara2017,Navin2017,Hagen2018,Kumar2021}. The interplay of different collective behaviors in this region makes it one of the most complex and intriguing areas of the nuclear landscape.

% Most studies in this region have focused on even-$Z$ nuclei, as they are easier to measure and interpret theoretically, but spectroscopic data on odd-$Z$ nuclei are essential for understanding the effect of unpaired nucleons on the composition of the nuclear wave function. Nb isotopes lie at the boundary between axially deformed Zr nuclei and triaxially deformed Mo isotopes, making them ideal candidates for investigating the onset of triaxiality in this region. Understanding the nuclear structure of neutron-rich Nb isotopes could also help clarify the evolution of nuclear shapes as a function of neutron number and shed light on the mechanisms driving shape coexistence.

The nuclear structure for odd-$A$ nuclei in this region is also important because they provide an understanding of the orbitals relevant to the large deformation and shape coexistence in this region. Most studies in this region have focused on even-$Z$ nuclei, as they are easier to measure and interpret theoretically. For example, experimental behavior of the low lying $\nu h_{11/2}~5/2^-[532]$ intruder bands in the odd-$A$ Zr isotopes~\cite{Hotchkis1991,Smith2012} provides substantial occupancy of the low $\Omega$ orbitals of the $\nu h_{11/2}$ intruder of the neighboring even-$A$ Zr core. Such occupancy was predicted by deformed mean field calculations and proposed as the driving force of large deformation in this region~\cite{Kumar1985,Mei2012}. Furthermore, such $\nu h_{11/2}$ bands also implies the triaxiality in Mo isotopes, while the Zr isotopes still have axially symmetry shapes~\cite{Hua2004}. On the other hand, spectroscopic data on odd-$Z$ nuclei are essential for understanding the effect of unpaired nucleons on the composition of the nuclear wave function, and the occupancy of the $\pi g_{9/2}$ orbitals around $Z=40$ spherical sub-shell closure. The latter one is related to the stability of the deformation~\cite{Hotchkis1991} and phase transition in this region~\cite{Togashi2016}. The Nb isotopes lie at the boundary between axially deformed Zr nuclei and triaxially deformed Mo isotopes, making them ideal candidates for investigating the onset of triaxiality in this region. Understanding the nuclear structure of neutron-rich Nb isotopes could also help clarify the evolution of nuclear shapes as a function of neutron number and could shed light on the mechanisms driving shape coexistence.

$^{99}$Nb has been reported to exhibit a weakly oblate ground state and deformed prolate excited states~\cite{Kumar2023}. In contrast, the isotopes $^{101,103,105,107}$Nb have been observed to possess triaxial prolate ground-state bands based on the $5/2^+[422]$ configuration~\cite{Luo2005,Hagen2017}. In 
$^{109}$Nb, an oblate-shaped isomeric state has been proposed at an excitation energy of 313~keV~\cite{Watanabe2011}. The most exotic Nb isotope observed to date is $^{113}$Nb, in which a $0.7~\mu $s isomer has been identified at 135~keV~\cite{Wu2022}. However, relatively limited information is available regarding the high-spin excited states in $^{105,107,109}$Nb. Further investigation into these levels could provide valuable insight into the evolution of nuclear shapes beyond the $N=60$ subshell, as well as the shape-driving effects of various orbitals.

In this work, we present a detailed spectroscopy study of odd-mass $^{105-109}$Nb isotopes using state-of-the-art high-resolution $\gamma$-ray spectroscopy techniques. A revised level scheme is also proposed for $^{99}$Nb. The experimental results are analyzed through systematic comparisons along the Nb isotopic chain and with neighboring isotones. Band assignments are based on systematics while the degree of triaxiality is discussed from the signature splitting. This approach follows the methodology established in previous studies~\cite{Luo2005}, and is here extended to more neutron-rich isotopes.

\section{Experimental details}
\label{sec:setup}

This work results from a combination of data from three different experiments: two $^{9}$Be($^{238}$U,f) induced fission experiments at the Grand Accélérateur National d'Ions Lourds (GANIL) facility and one $^{252}$Cf spontaneous fission experiment at the Lawrence Berkeley National Laboratory (LBNL). This complementarity allowed the investigation of the neutron-rich Nb isotopes using both the unambiguous isotopic identification of the fission fragments using the VAMOS++ spectrometer and the high-fold data from Gammasphere. Such a combined analysis has already proven to be highly effective in previous studies on Pr, Pm, and Y isotopes using EXOGAM and Gammasphere data~\cite{Wang2015,Bhattacharyya2018,Wang2021}.

The two GANIL experiments were conducted as part of the AGATA fission campaign~\cite{Lemasson2023} and are designated as E661 and E680. Fission fragments were produced via transfer and fusion-induced fission reactions, using a 6.2~MeV/u $^{238}$U beam at a typical intensity of $\sim 1$~pnA, impinging on $^{9}$Be targets with thicknesses of 1.6, 5 and 10~$\mu$m. Both experiments employed a similar setup, combining the large-acceptance VAMOS++ magnetic spectrometer~\cite{Rejmund_2011} with the AGATA $\gamma$-ray tracking array~\cite{Akkoyun2012,Clement2017}. 

The AGATA array was used in a compact configuration (13.5~cm from the target) to detect prompt $\gamma$ rays. The large segmentation of the AGATA crystals, combined with a pulse-shape analysis technique~\cite{Bruyneel_2016,Venturelli_2004}, enabled precise localization of $\gamma$-ray interaction points. The $\gamma$-ray trajectory in the HPGe array was reconstructed using a tracking algorithm~\cite{Lopez-Martens_2004} to obtain the total $\gamma$-ray energy along with the position of its first interaction. By combining the velocity vector measurement of fission fragments ($v/c \approx 0.1$) in VAMOS++ with the determination of the first $\gamma$-ray interaction position in AGATA, a precise event-by-event Doppler correction was applied, achieving a $\gamma$-ray energy resolution of 5~keV (FWHM) at 1.2 MeV. Further details on the experimental setup and characterization can be found in~\cite{Rejmund_2011,Navin_2014,Lemasson2023,Kim2017, Abushawish2024}.

\begin{figure}
\includegraphics[width=\linewidth]{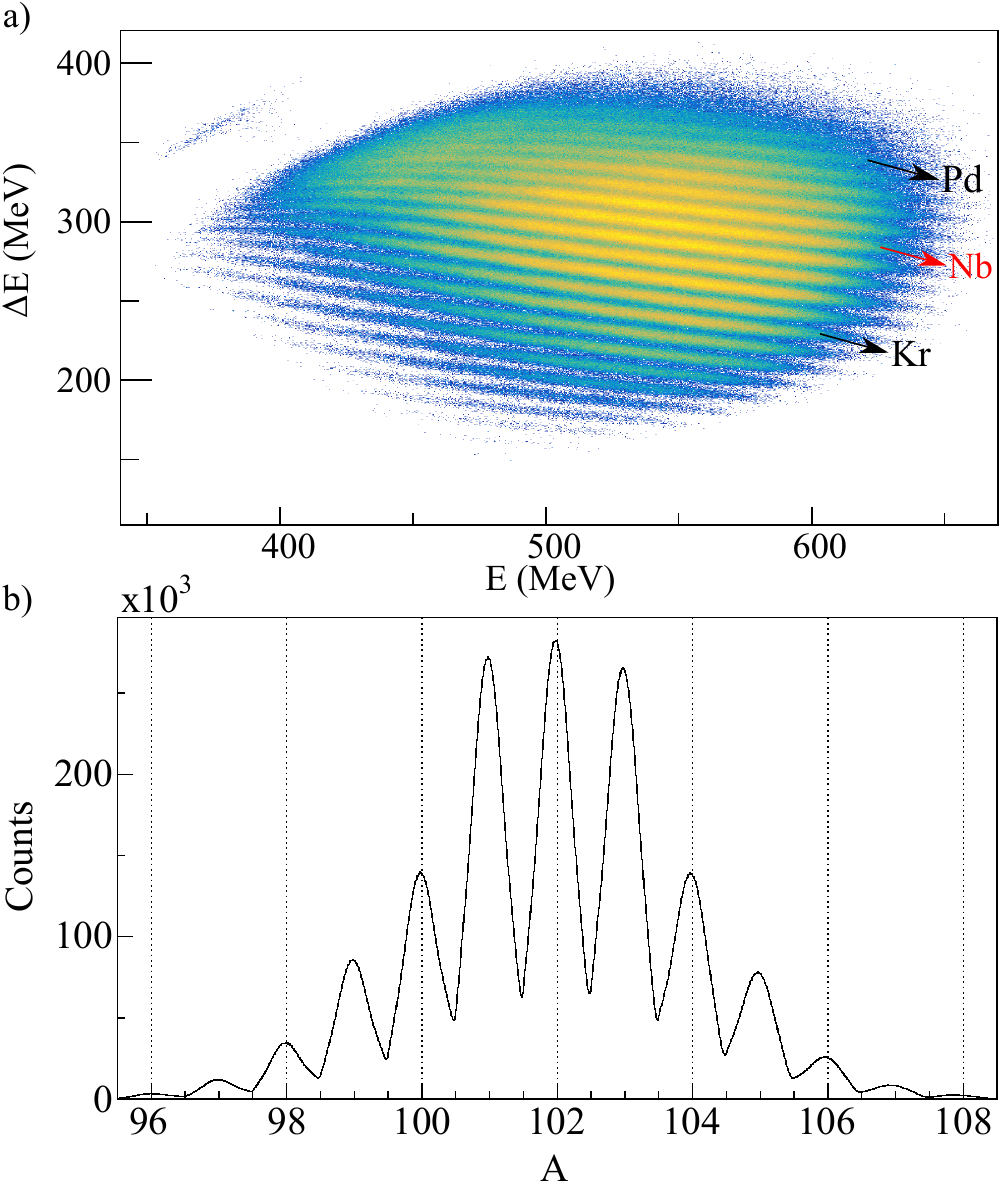}% 
\caption{VAMOS++ identification spectra for fission fragments produced in the E680 experiment. (a) Two-dimensional spectrum of energy loss ($\Delta$E) as a function of total energy (E) measured in ionization chambers of VAMOS++. The elements Kr, Nb and Pd are labeled. (b) A distribution for $Z=41$ isotopes identified in VAMOS++ measured in coincidence with $\gamma$ rays in AGATA.}
\label{fig:Z-A-identification}
\end{figure}

The E680 experiment focused on studying nuclei near $^{78}$Ni. The VAMOS++ angle was consequently set to 28° to maximize acceptance on the lightest fission fragments. The VAMOS++ identification spectra for the E680 experiment are shown in Fig.~\ref{fig:Z-A-identification}. The atomic number ($Z$) was determined from the $\Delta$E-E analysis (a), while the atomic mass ($A$) was extracted by combining total energy and velocity measurements. The mass distribution of Nb isotopes is shown in panel Fig.~\ref{fig:Z-A-identification} (b).

The E661 experiment focused on the heavier fission fragments, with a VAMOS++ angle set at 20°. However, as shown in the identification spectra of this experiment (Fig.~1 of Ref.~\cite{Lemasson2023}), Nb nuclei were still produced in statistically significant amounts thanks to the large acceptance in $B\rho$. Additionally, for this experiment, seven EXOGAM HPGe Clover detectors~\cite{Simpson2000} were arranged in a wall-like configuration at the focal plane of the VAMOS++ spectrometer to detect delayed $\gamma$ rays ($\gamma_D$) within a time window ranging from 100~ns to 200~$\mu$s. Details on this setup and the analysis methods are discussed in Ref.~\cite{Kim2017}.

The LBNL experiment used a 62~$\mu$Ci spontaneous $^{252}$Cf fission source. The Gammasphere array, composed of 101 HPGe detectors, enabled the measurement of high-fold $\gamma$ events with high statistics ($5.7\times 10^{11}$ $\gamma-\gamma-\gamma$ in 3D cubes and $1.9\times 10^{11}$ $\gamma-\gamma-\gamma-\gamma$ in 4D hypercubes). By applying various coincidence time windows, state lifetimes ranging from 2~ns to a few hundred ns were measured. Further experimental details can be found in Ref.~\cite{Hwang2006, Wang2015_PhD}.

The analysis first used the Z- and A-gated AGATA data to obtain $\gamma$-ray spectra unambiguously attributed to their respective nuclei. This was followed by a high-fold $\gamma$-ray analysis using the Gammasphere data with the RADWARE software package~\cite{Radford1995}.

AGATA data were used for extracting the energy and relative intensities of the $\gamma$-ray presented in this work. The experimental uncertainties on the $\gamma$-ray energies result from the combination of the statistical error, obtained from peak fitting, and systematic errors. The latter were determined by comparing literature values with experimental values for approximately 50 of the most intense transitions observed in this dataset. A systematic uncertainty of 0.2~keV estimated from this analysis was primarily due to uncertainties in the kinematic reconstruction of the fission fragment trajectories. The reported relative intensities for the different transitions were corrected for the detector efficiencies, with uncertainties including both statistical and systematic errors (5\%, estimated from GEANT4 simulations). AGATA data were analyzed using the Cubix software~\cite{cubix}.

Although AGATA is in principle capable of performing angular-correlation measurements~\cite{Lauritsen2025}, the limited number of operating crystals for the present experiments did not provide sufficient angular coverage to extract reliable correlation coefficients.

\section{Results}\label{sec:results}

\begin{figure}
\includegraphics[width=\linewidth]{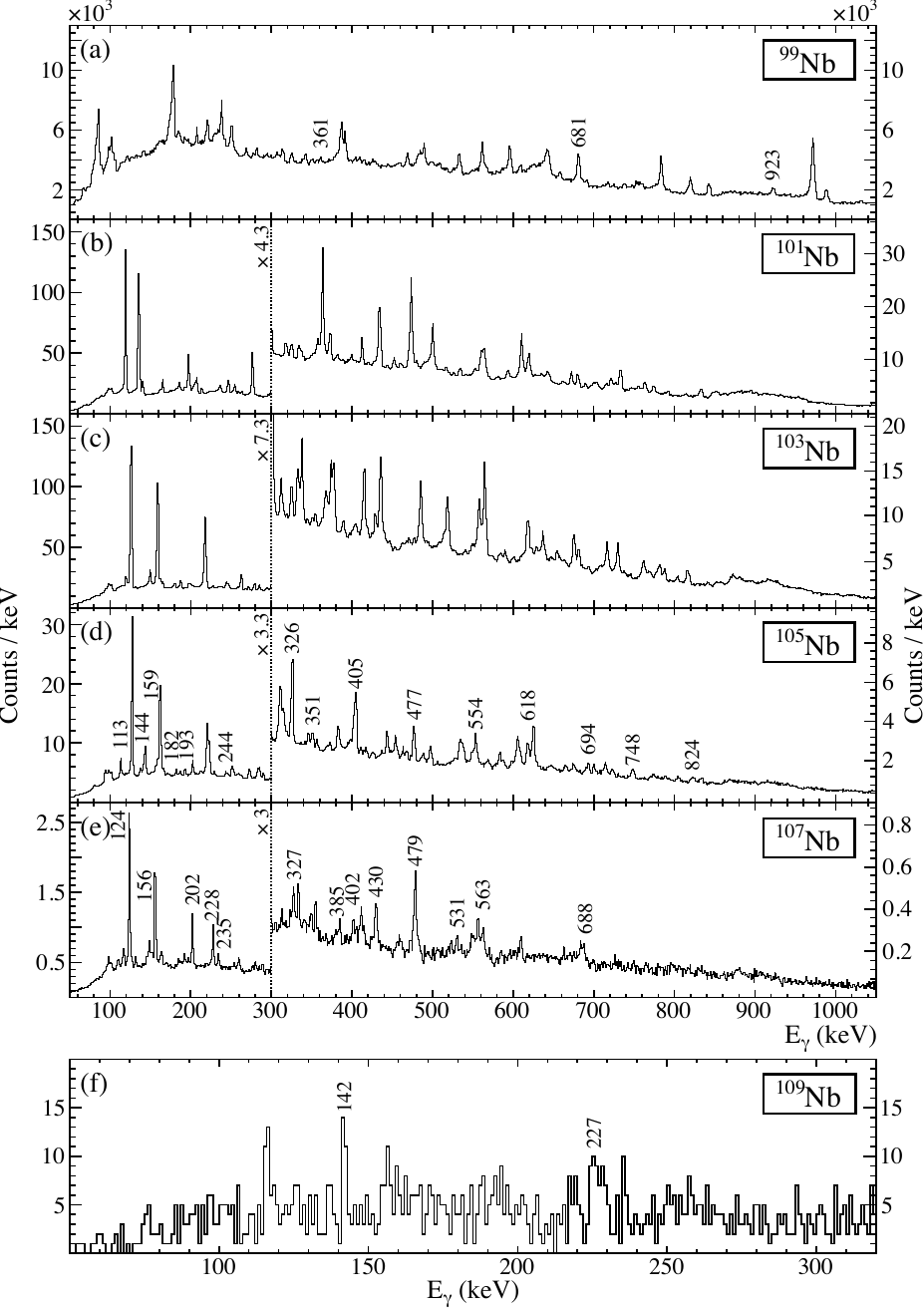}% 
\caption{Doppler-corrected $\gamma$-ray spectra measured in AGATA of $^{99,101,103,105,107,109}$Nb isotopically identified with VAMOS++. The labeled transitions are new. For $\gamma$-ray energies above 300~keV, the Y-axis is displayed using a different scale when necessary, associated with the right Y-axis (in Counts/keV).}
\label{fig:singles}
\end{figure}

This section presents the new transitions measured in $^{99,105,107,109}$Nb along with their associated level schemes. Each nucleus is described in a dedicated subsection. Spin assignments and bands interpretation are briefly introduced here but are discussed in more detail in Section~\ref{sec:interpretation}.

The tracked AGATA singles spectra, isotopically identified using VAMOS++, are shown in Fig.~\ref{fig:singles} for $^{99, 101, 103, 105 ,107, 109}$Nb. The labeled transitions correspond to those reported for the first time in this work. A clear evolution in the overall spectral shape is observed between $^{99}$Nb and the heavier Nb isotopes, highlighting the shape transition occurring in this region between $N=58$ and $N=60$. As can be seen from the figure, for $^{101-109}$Nb ($N\ge60$), the spectra are dominated by low-energy transitions, indicative of the highly deformed nature of these nuclei.

\subsection{\texorpdfstring{$^{99}$}{99}Nb}

The structure of $^{99}$Nb was first established from $\beta$-decay studies of $^{99}$Zr~\cite{Ohm1989,Pfeiffer1984,Lhersonneau1998}. Lifetime measurements of several states were performed~\cite{Ohm1989,Pfeil2023} and a long-lived isomer at 365~keV with a lifetime of 2.5~minutes was identified. Because of the low spin of the parent nucleus ($1/2^+$), high-spin states could not be populated in $\beta$-decay. However, a recent study~\cite{Kumar2023} combining fusion-fission and multinucleon grazing reactions successfully populated high-spin states of the $^{99}$Nb, revealing two distinct bands: a negative-parity band based on the long-lived $1/2^-$ isomer and a positive-parity ground-state band. Both bands are observed in the present work; however, a revised level scheme is proposed for the positive-parity band.

\begin{table}
\caption{Level energies ($E_i$), $\gamma$-ray energies (E$_\gamma$), and relative intensities (I$_\gamma$) of transitions assigned to $^{99}$Nb in this work, along with the spin and parity of the initial ($J_i$) and final ($J_f$) states. Intensities are given relative to the strongest observed transition. New levels and transitions are in bold. For clarity, the two bands are listed separately in the table. A dash symbol indicates that no spin assignment is proposed. The excitation energy of the $1/2^-$ isomer has been taken from~\cite{NDS_A99}.}
\begin{ruledtabular}
\begin{tabular}{lllcc}
$E_i$ (keV)         & E$_\gamma$ (keV)      & I$_\gamma$ $(\%)$ & $J_i^\pi$             &  $J_f^\pi$            \\
\hline     
0                   &                       &                   & $9/2^+$               &                       \\
971.9(2)            & 971.9(2)              & 100               & $(13/2^+)$            & $9/2^+$               \\
\textbf{1653.1(4)}  & \textbf{681.2(3)}     & \textbf{32(4)}    & \textbf{-}            & $\mathbf{(13/2^+)}$   \\
1755.6(4)           & 783.7(3)              & 39(6)             & $(17/2^+)$            & $(13/2^+)$            \\
\textbf{2575.7(4)}  & 819.9(3)              & 26(3)             & $\mathbf{(21/2^+)}$   & $\mathbf{(17/2^+)}$   \\
                    & \textbf{922.7(3)}     & \textbf{13(2)}    & $\mathbf{(21/2^+)}$   & \textbf{-}            \\
\textbf{3419.0(4)}  & 843.3(3)              & 16(3)             & $\mathbf{(25/2^+)}$   & $\mathbf{(21/2^+)}$   \\
3810.5(5)           & 391.5(3)              & 22.2(19)          & -                     & $\mathbf{(25/2^+)}$   \\
\textbf{4171.7(6)}  & \textbf{361.2(5)}     & \textbf{2.7(9)}   & \textbf{-}            & -                     \\
\multicolumn{5}{l}{\vspace{-8pt}} \\
\hline
\multicolumn{5}{l}{\vspace{-8pt}} \\
365.27(8)           &                       &                   & $1/2^-$               &                       \\
543.6(4)            & 178.3(3)              & 52(8)             & $3/2^-$               & $1/2^-$               \\
629.6(5)            & 86.0(3)               & 61(9)             & $5/2^-$               & $3/2^-$               \\
868.8(4)            & 238.6(3)              & 24(3)             & $(7/2^-)$             & $5/2^-$               \\
                    & 325.8(4)              & 6.1(9)            & $(7/2^-)$             & $3/2^-$               \\
1119.6(4)           & 251.1(3)              & 12(2)             & $(9/2^-)$             & $(7/2^-)$             \\
                    & 489.6(3)              & 24(2)             & $(9/2^-)$             & $5/2^-$               \\
1402.0(3)           & 533.5(3)              & 22(3)             & $(11/2^-)$            & $(7/2^-)$             \\
                    & 282.1(3)              & 5.2(10)           & $(11/2^-)$            & $(9/2^-)$             \\
1716.1(3)           & 596.0(3)              & 31(3)             & $(13/2^-)$            & $(9/2^-)$             \\
                    & 314.4(3)              & 9.5(11)           & $(13/2^-)$            & $(11/2^-)$            \\
2043.6(4)           & 641.5(5)              & 18(4)             & $(15/2^-)$            & $(11/2^-)$            \\
                    & 328(2)                & <1                & $(15/2^-)$            & $(13/2^-)$            \\
2360.2(3)           & 644.1(5)              & 19(5)             & $(17/2^-)$            & $(13/2^-)$            \\
\end{tabular}
\end{ruledtabular}
\label{tab:99nb}
\end{table}

\begin{figure}
\includegraphics[width=\linewidth]{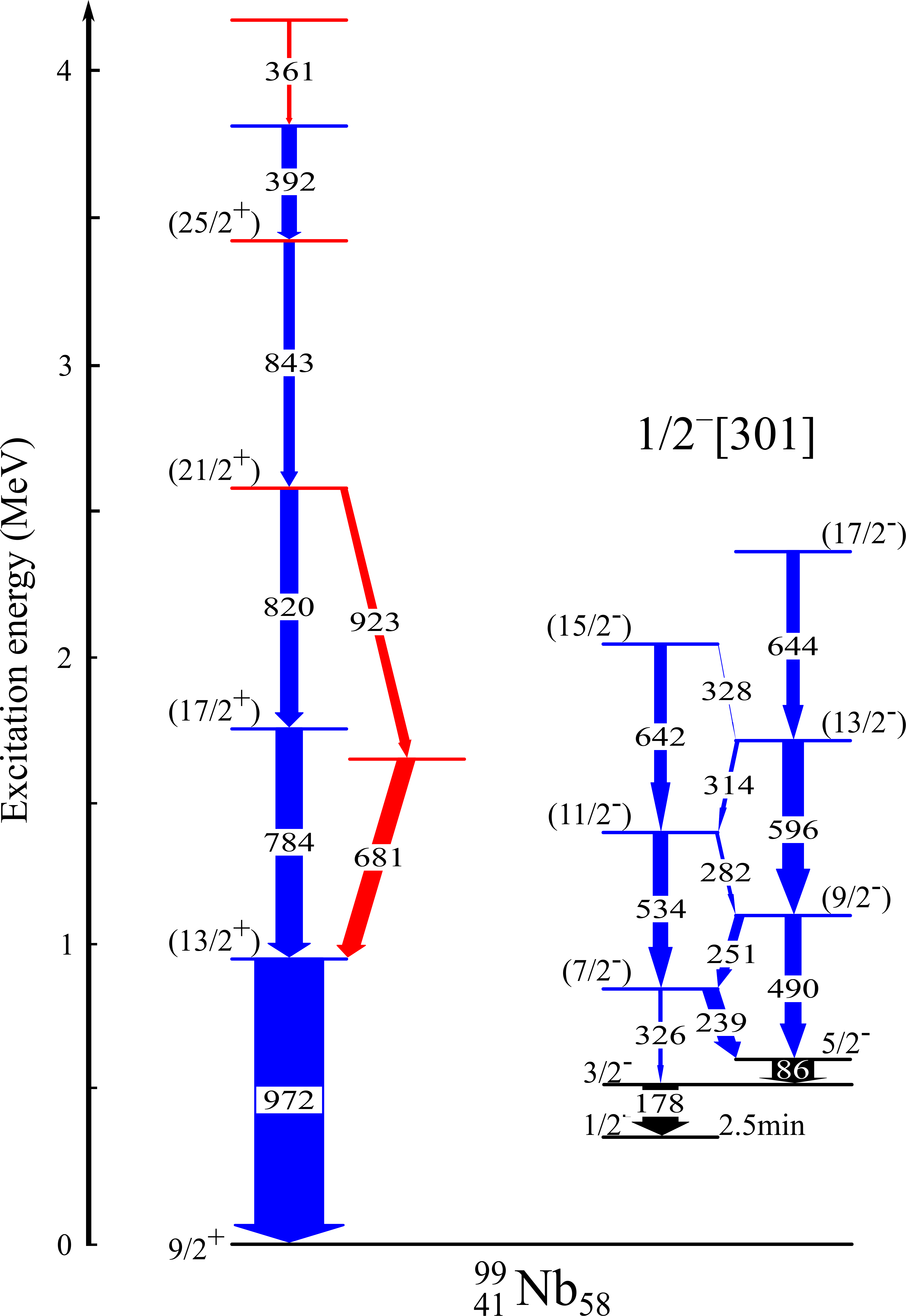}% 
\caption{Revised level scheme of $^{99}$Nb. New transitions and levels are marked in red, while those in blue correspond to those recently reported~\cite{Kumar2023}. The width of the arrows represents the observed intensities relative to the strongest transition.}
\label{fig:99Nb_LS}
\end{figure}

The $\gamma$-ray spectrum of $^{99}$Nb nuclei, $Z$ and $A$ identified using VAMOS++, is shown in Fig.~\ref{fig:singles} (a). The labeled transitions are reported here for the first time. The $\gamma$-ray transitions observed in this work along with their relative intensities are listed in Table~\ref{tab:99nb}. Most of the levels populated via $\beta$-decay are non-yrast and are not observed in this work, except for the 178 and 86~keV transitions. Three new transitions, with energies of 361, 681, and 923~keV, are in bold. The updated level scheme is shown in Fig.~\ref{fig:99Nb_LS}, where transitions and states recently published in~\cite{Kumar2023} are shown in blue, while new results from this work are shown in red. Although the negative-parity band is confirmed, a revised structure is proposed for the positive-parity band, featuring a different ordering of the ground-state band transitions. The 392~keV transition was identified in~\cite{Kumar2023} as depopulating the $(21/2^+)$ state to the $(17/2^+)$, while in this work, it is proposed on the top of the band, populating the $(25/2^+)$ state.

\begin{figure}
\includegraphics[width=\linewidth]{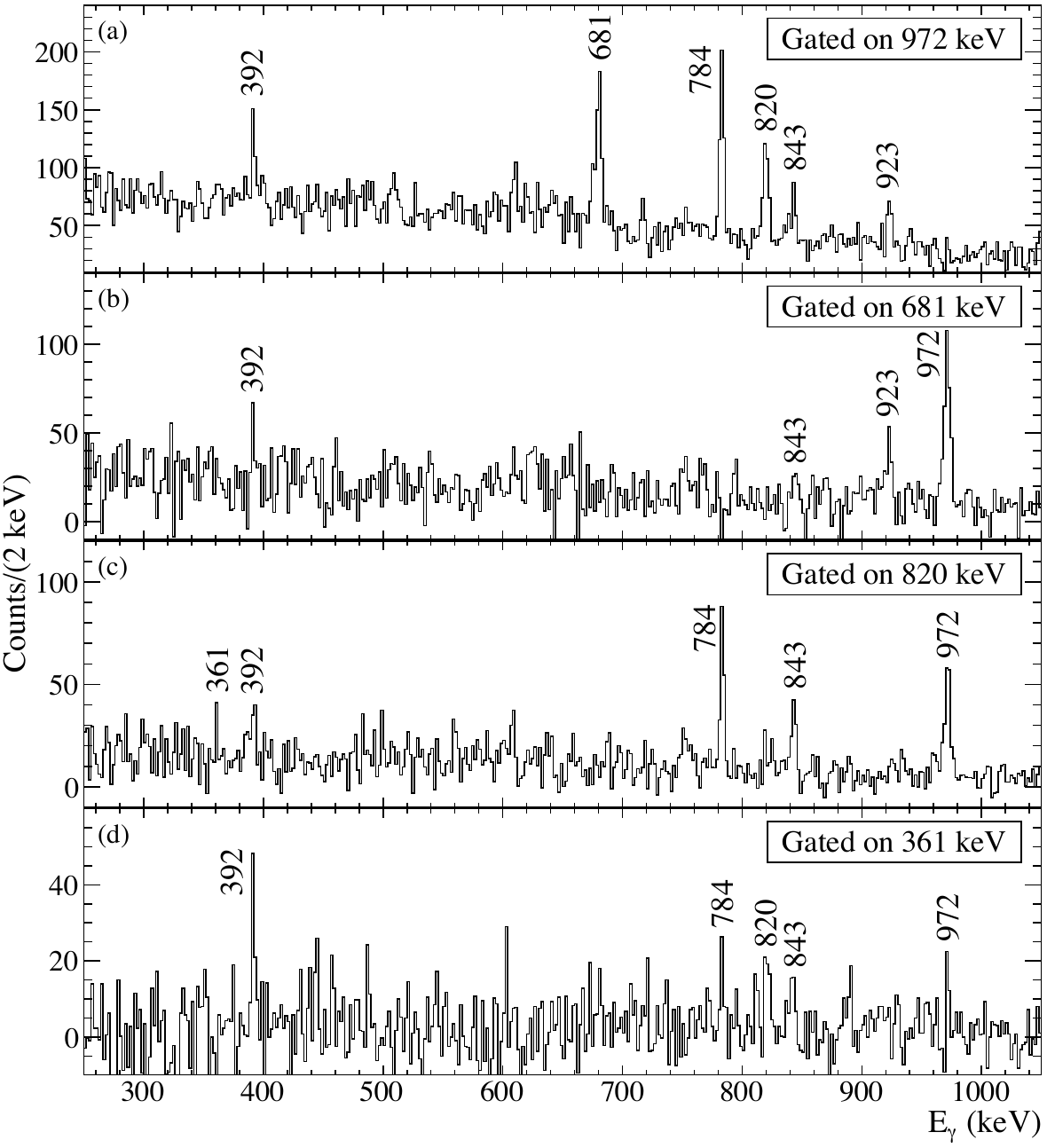}% 
\caption{Doppler-corrected $\gamma$-ray coincidence spectra for selected gates in the positive-parity band of $^{99}$Nb shown in Fig.~\ref{fig:99Nb_LS}.}
\label{fig:99Nb_gates}
\end{figure}

An example of coincidence spectra is shown in Fig.~\ref{fig:99Nb_gates}, confirming the new level scheme proposed for the positive-parity band of $^{99}$Nb. The spectrum gated on the 972~keV transition (a) shows a strong coincidence with the 681~keV and 923~keV $\gamma$ rays, which were not reported in~\cite{Kumar2023}. The 681 and 923~keV transitions are observed in coincidence with the main band, except for the 784 and 820~keV $\gamma$ rays (b). The same applies to the transitions at 784 and 820~keV, which are observed with the entire band except for the 681 and 923~keV $\gamma$ rays (c). This suggests that these two pairs of $\gamma$ rays are placed in parallel in the level scheme. Since their summed energy is identical, this supports their proposed placement in the level scheme. The spectrum gated on the 361~keV transition (d) shows coincidences with the entire ground-state band. As this is the $\gamma$ ray with the lowest intensity, we propose placing it at the top of the band. The ordering of the $\gamma$ rays is based on their relative intensity ratio. The 843 and 392~keV $\gamma$ rays have comparable intensities in the singles spectrum, but comparisons of the gated intensities, such as in the spectrum gated on the 820~keV transition (c), confirm the proposed transition ordering: after efficiency correction, the 843~keV transition is found to be approximately 35\% more intense than the 392~keV one. A detailed interpretation of the nature of these two structures is provided in Section~\ref{sec:interpretation}.

\subsection{\texorpdfstring{$^{101,103}$}{101-103}Nb}

It is worth noting that in the datasets used in this work, $^{101}$Nb and $^{103}$Nb were produced with high statistics, as shown in Fig.~\ref{fig:singles} (b) and (c) respectively. Nevertheless, no new transitions were observed, hence are not included in this systematic study of neutron-rich odd-even Nb isotopes. The most detailed spectroscopy of $^{101}$Nb was obtained from $^{252}$Cf fission data at Gammasphere~\cite{Luo2005}, while the spectroscopy of $^{103}$Nb was derived from the $^{238}$U$(\alpha,f)$ fusion-fission reaction with the Gammasphere and CHICO detector arrays~\cite{Hua2002}.

\subsection{\texorpdfstring{$^{105}$}{105}Nb}
\label{sec:105Nb}

The ground-state band of $^{105}$Nb was initially identified in~\cite{Hotchkis1991} and later extended in~\cite{Luo2005}. One- and two-phonon $\gamma$-vibrational bands were subsequently identified in Ref.~\cite{Li2013}. In this work, 18 new transitions associated with $^{105}$Nb were identified. The $\gamma$-ray spectrum of $^{105}$Nb nuclei is shown in Fig.~\ref{fig:singles} (d), with the newly observed transitions labeled. The $\gamma$-ray transitions observed in this work, along with their relative intensities, are listed in Table~\ref{tab:105Nb}. New levels and transitions are in bold. 

\begin{figure*}
\includegraphics[width=\linewidth]{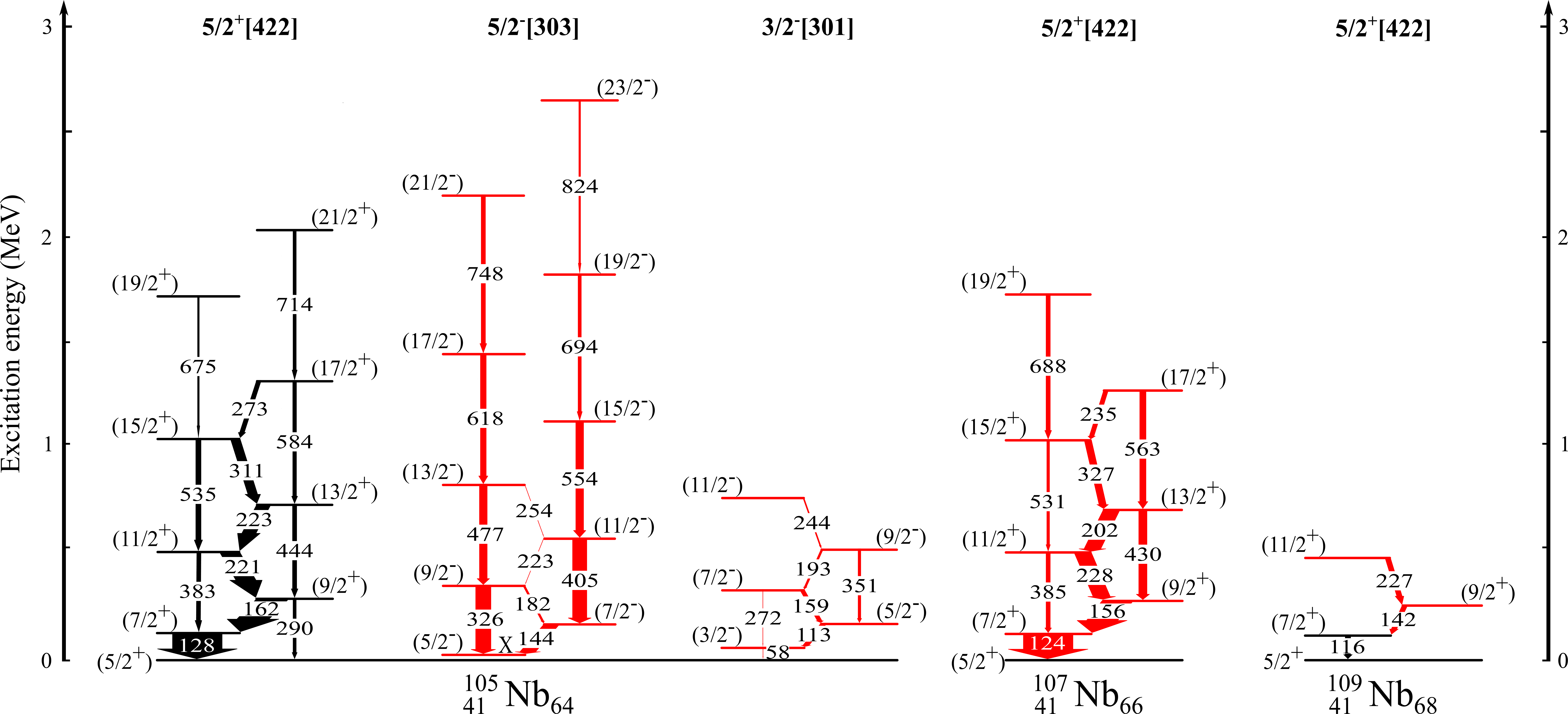}% 
\caption{Level schemes of $^{105, 107, 109}$Nb obtained from the present work. New $\gamma$-ray transitions and levels are marked in red. The label “X” indicates unknown band-head energy.}
\label{fig:105_107Nb_LS}
\end{figure*}

\begin{table}[!h]
\caption{Level energies ($E_i$), $\gamma$-ray energies (E$_\gamma$), and relative intensities (I$_\gamma$) of transitions assigned to $^{105}$Nb in this work, along with the spin and parity of the initial ($J_i$) and final ($J_f$) states. Intensities are given relative to the strongest observed transition. New levels and transitions are in bold. A label ($^{g}$) is added to transitions observed exclusively with Gammasphere. X denotes the unknown energy of the band-head corresponding to the $\pi 5/2^-[303]$ Nilsson orbital. For clarity, the three bands are listed separately in the table. The dash symbol indicates that no relative intensity value could be obtained.}
\begin{ruledtabular}
\begin{tabular}{lllcc}
$E_i$ (keV)             & E$_\gamma$ (keV)      & I$_\gamma$ $(\%)$     & $J_i^\pi$             &  $J_f^\pi$                \\
\hline
0                       &                       &                       & $(5/2^+)$             &                           \\
127.6(2)                & 127.6(2)              & 100                   & $(7/2^+)$             & $(5/2^+)$                 \\
289.8(3)                & 162.2(2)              & 66(5)                 & $(9/2^+)$             & $(7/2^+)$                 \\
                        & 289.8(3)              & 7.1(7)                & $(9/2^+)$             & $(5/2^+)$                 \\
510.5(3)                & 220.6(2)              & 41(4)                 & $(11/2^+)$            & $(9/2^+)$                 \\
                        & 383.2(3)              & 8.8(12)               & $(11/2^+)$            & $(7/2^+)$                 \\
733.8(2)                & 223.3(2)              & 29(3)                 & $(13/2^+)$            & $(11/2^+)$                \\
                        & 443.9(3)              & 8.9(9)                & $(13/2^+)$            & $(9/2^+)$                 \\
1044.8(2)               & 310.9(3)              & 19.9(16)              & $(15/2^+)$            & $(13/2^+)$                \\
                        & 534.5(3)              & 10.5(14)              & $(15/2^+)$            & $(11/2^+)$                \\
1317.5(2)               & 272.8(3)              & 9.4(11)               & $(17/2^+)$            & $(15/2^+)$                \\
                        & 583.5(3)              & 8.2(13)               & $(17/2^+)$            & $(13/2^+)$                \\
1720.0(3)               & 675.2(4)              & 4.4(8)                & $(19/2^+)$            & $(15/2^+)$                \\
2031.8(2)               & 714.3(3)              & 7.1(10)               & $(21/2^+)$            & $(17/2^+)$                \\
\multicolumn{5}{l}{\vspace{-8pt}} \\ 
\hline
\multicolumn{5}{l}{\vspace{-8pt}} \\
\textbf{X+0}            &                       &                       & $\mathbf{(5/2^-_1)}$  &                          \\
\textbf{X+143.7(2)}     & \textbf{143.7(2)}     & \textbf{32(3)}        & $\mathbf{(7/2^-_1)}$  & $\mathbf{(5/2^-_1)}$     \\
\textbf{X+326.0(2)}     & \textbf{326.1(2)}     & \textbf{31(3)}        & $\mathbf{(9/2^-_1)}$  & $\mathbf{(5/2^-_1)}$     \\
                        & \textbf{182.1(3)}     & \textbf{4.9(7)}       & $\mathbf{(9/2^-_1)}$  & $\mathbf{(7/2^-_1)}$     \\
\textbf{X+548.3(3)}     & \textbf{404.5(3)}     & \textbf{29(4)}        & $\mathbf{(11/2^-_1)}$ & $\mathbf{(7/2^-_1)}$     \\
                        &\textbf{222.6(6)}$^{g}$& \textbf{-}           & $\mathbf{(11/2^-_1)}$ & $\mathbf{(9/2^-_1)}$     \\
\textbf{X+802.8(3)}     & \textbf{476.9(3)}     & \textbf{15.5(13)}     & $\mathbf{(13/2^-)}$   & $\mathbf{(9/2^-_1)}$     \\
                        & \textbf{254(1)}$^{g}$ & \textbf{-}           & $\mathbf{(13/2^-)}$   & $\mathbf{(11/2^-_1)}$    \\
\textbf{X+1102.0(4)}    & \textbf{553.6(3)}     & \textbf{16.3(14)}     & $\mathbf{(15/2^-)}$   & $\mathbf{(11/2^-_1)}$    \\
\textbf{X+1421.0(4)}    & \textbf{618.2(3)}     & \textbf{12.3(13)}     & $\mathbf{(17/2^-)}$   & $\mathbf{(13/2^-)}$      \\
\textbf{X+1795.5(4)}    & \textbf{693.5(3)}     & \textbf{7.3(10)}      & $\mathbf{(19/2^-)}$   & $\mathbf{(15/2^-)}$      \\
\textbf{X+2169.4(4)}    & \textbf{748.4(3)}     & \textbf{8.9(11)}      & $\mathbf{(21/2^-)}$   & $\mathbf{(17/2^-)}$      \\
\textbf{X+2619.6(6)}    & \textbf{824.1(6)}     & \textbf{4.0(9)}       & $\mathbf{(23/2^-)}$   & $\mathbf{(19/2^-)}$      \\
\multicolumn{5}{l}{\vspace{-8pt}} \\
\hline
\multicolumn{5}{l}{\vspace{-8pt}} \\
\textbf{58.1(1)}        & \textbf{58.1(1)}$^{g}$& \textbf{-}            & $\mathbf{(3/2^-)}$  & $\mathbf{(5/2^+)}$       \\
\textbf{171.3(3)}       & \textbf{113.2(2)}     & \textbf{12.3(12)}     & $\mathbf{(5/2^-_2)}$  & $\mathbf{(3/2^-)}$     \\
\textbf{330.2(4)}       & \textbf{158.9(3)}     & \textbf{10.1(9)}      & $\mathbf{(7/2^-_2)}$  & $\mathbf{(5/2^-_2)}$     \\
                        & \textbf{272(1)}       & \textbf{<4}           & $\mathbf{(7/2^-_2)}$  & $\mathbf{(3/2^-)}$     \\
\textbf{522.8(3)}       & \textbf{351.2(3)}     & \textbf{4.8(12)}      & $\mathbf{(9/2^-_2)}$  & $\mathbf{(5/2^-_2)}$     \\
                        & \textbf{193.0(3)}     & \textbf{5.1(14)}      & $\mathbf{(9/2^-_2)}$  & $\mathbf{(7/2^-_2)}$     \\
\textbf{766.3(4)}       & \textbf{243.5(4)}     & \textbf{2.8(3)}       & $\mathbf{(11/2^-_2)}$ & $\mathbf{(9/2^-_2)}$     \\
\end{tabular}
\end{ruledtabular}
\label{tab:105Nb}
\end{table}

\begin{figure}
\includegraphics[width=\linewidth]{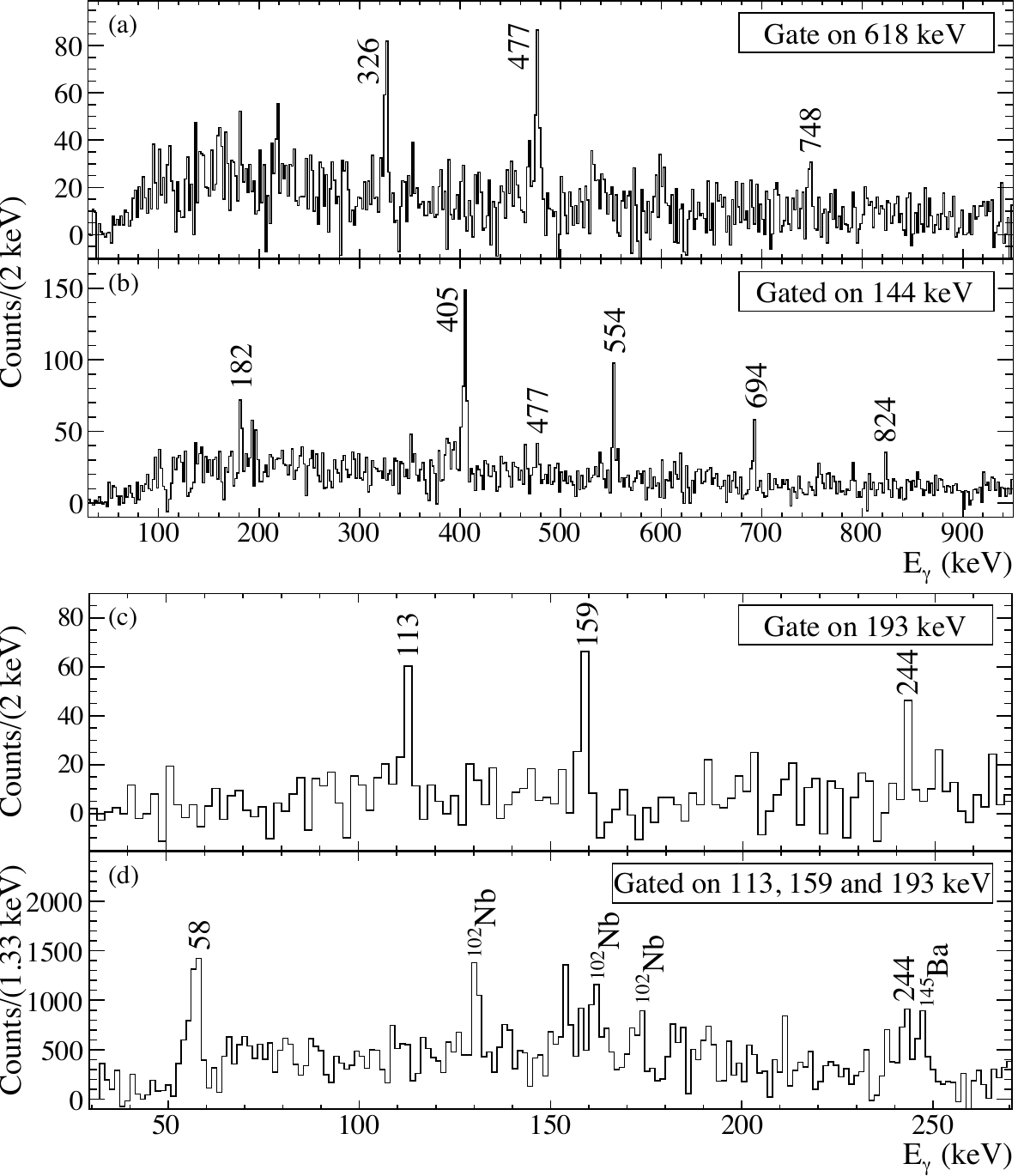}% 
\caption{Coincidence spectra for selected transitions in $^{105}$Nb: a,b) AGATA spectra gated on the 618 and 144~keV transitions ($\pi 5/2^-[303]$ band). c) AGATA spectrum gated on the 193~keV transition ($\pi 3/2^-[301]$ band). d) Gammasphere triple-coincidence spectrum gated on the 113, 159, and 193~keV transitions ($\pi 3/2^-[301]$ band). The corresponding bands are shown in Fig.~\ref{fig:105_107Nb_LS}.}
\label{fig:105Nb_gates}
\end{figure}

The new level scheme is presented in Fig.~\ref{fig:105_107Nb_LS}, with new transitions and states marked in red. We confirm the ground-state band associated with the $5/2^+[422]$ Nilsson orbital as reported in~\cite{Luo2005} and identify two new bands for the first time. An example of coincidence spectra is shown in Fig.~\ref{fig:105Nb_gates}.

The first band consists of two interconnected cascades: one with the 326, 477, 618, and 748~keV transitions, and the other with the 405, 554, 694, and 824~keV transitions, as shown in the coincidence spectra (a) and (b) of Fig.~\ref{fig:105Nb_gates}, respectively. These two cascades are connected via transitions of significantly lower intensity, at 144, 182, 223, and 254~keV. Based on systematic comparisons with $^{101,103}$Nb~\cite{Luo2005,Hua2002} (discussed in Section~\ref{sec:interpretation}), this band is assigned to the $\pi 5/2^-[303]$ Nilsson orbital. The spin assignment is consequently proposed based on the systematics of $^{101,103}$Nb, as two E2 cascades linked through M1/E2 transitions, with a $(5/2^-)$ band head. While the $\gamma$-ray transition from the $(5/2^-)$ band head to the ground state is well populated for $^{101,103}$Nb in the present data, it is not the case for $^{105}$Nb.

The second newly identified band also consists of two interconnected cascades, with lower intensity compared to the $5/2^+[422]$ and $\pi 5/2^-[303]$ bands. Spectrum (c) in Fig.~\ref{fig:105Nb_gates} presents an example of a coincidence spectrum obtained with AGATA gated on the 193~keV $\gamma$-ray. As with the $\pi 5/2^-[303]$ band, the systematics of $^{101,103}$Nb have been used to assign the tentative spin and parity of the new levels, resulting in two E2 cascades linked through M1/E2 transitions, with a $(3/2^-)$ band head corresponding to the $\pi 3/2^-[301]$ Nilsson orbital. The $\gamma$-ray transition from the $(3/2^-)$ band head to the ground state, with an energy of 58.1(1)~keV, was detected using the Gammasphere data, as shown in the triple-gated spectrum in Fig.~\ref{fig:105Nb_gates} (d). A left-sided tail is observed, suggesting the presence of a doublet, with a weaker transition at 55.1(1)keV. 

% This $\gamma$-ray could correspond to a transition from the $\pi 3/2^-[301]$ to the $\pi 5/2^-[303]$ band-heads. Such a hypothesis would imply that the $\pi 5/2^-[303]$ band-head lies at an energy of 3~keV. However, the current experimental evidence is not sufficiently robust to confidently assign this energy to the $(5/2^-)$ state.

\begin{figure}
\includegraphics[width=\linewidth]{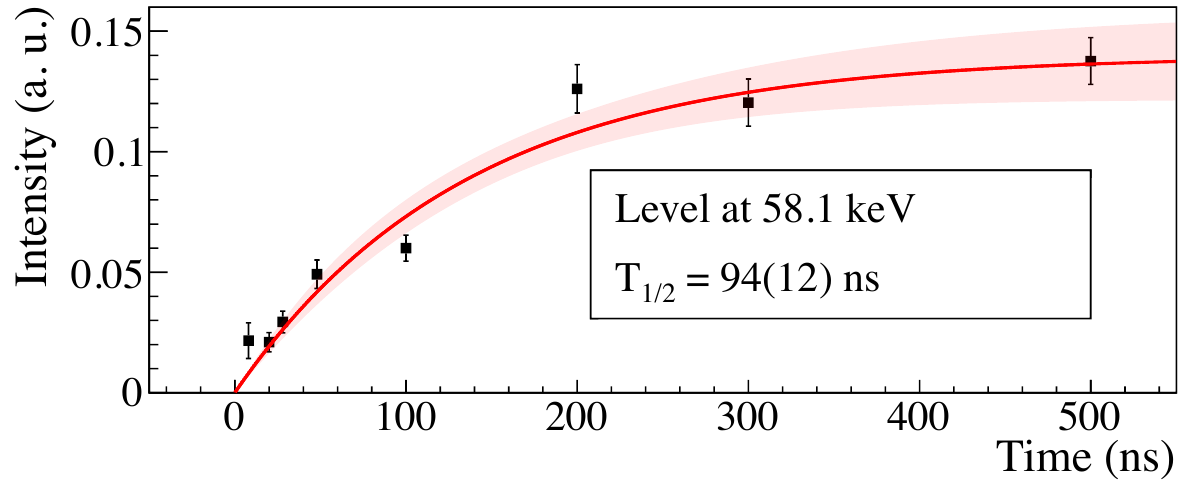}% 
\caption{Half-life measurement of the 58.1~keV $(3/2^-)$ state in $^{105}$Nb obtained using Gammasphere. The shaded area indicates the uncertainty of the fit.}
\label{fig:105Nb_58LT}
\end{figure}

The half-life of the $(3/2^-)$ state was determined from Gammasphere data to be $T_{1/2}=94(12)$~ns, as shown in Fig.~\ref{fig:105Nb_58LT}. The figure displays the transition intensity as a function of the time window used in the event builder. The shaded area represents the fit uncertainty. The error bars correspond to $2\sigma$ uncertainties, accounting for possible effects of a small 59~keV contamination, most likely originating from tungsten K$\alpha_1$ X-rays (59.3~keV). Further details on the lifetime evaluation process can be found in Ref.~\cite{Hwang2006}.

\subsection{\texorpdfstring{$^{107}$}{107}Nb}
\label{sec:107Nb}

The only published spectroscopy of $^{107}$Nb was obtained from $^{238}$U+$^{9}$Be transfer and fusion-induced fission reactions, using the EXOGAM and VAMOS++ spectrometers~\cite{Hagen2017}, in an experiment focused on lifetime measurements using a plunger device. Unlike the present work, the previous experiment suffered from limited statistics for $^{107}$Nb, which prevented the use of $\gamma$-$\gamma$ coincidence analysis. Additionally, the presence of material in the fragment path, required for lifetime measurements, degraded the $\gamma$-ray energy resolution and isotopic selectivity. As a result, the proposed level scheme for $^{107}$Nb remained tentative.

\begin{figure}
\includegraphics[width=\linewidth]{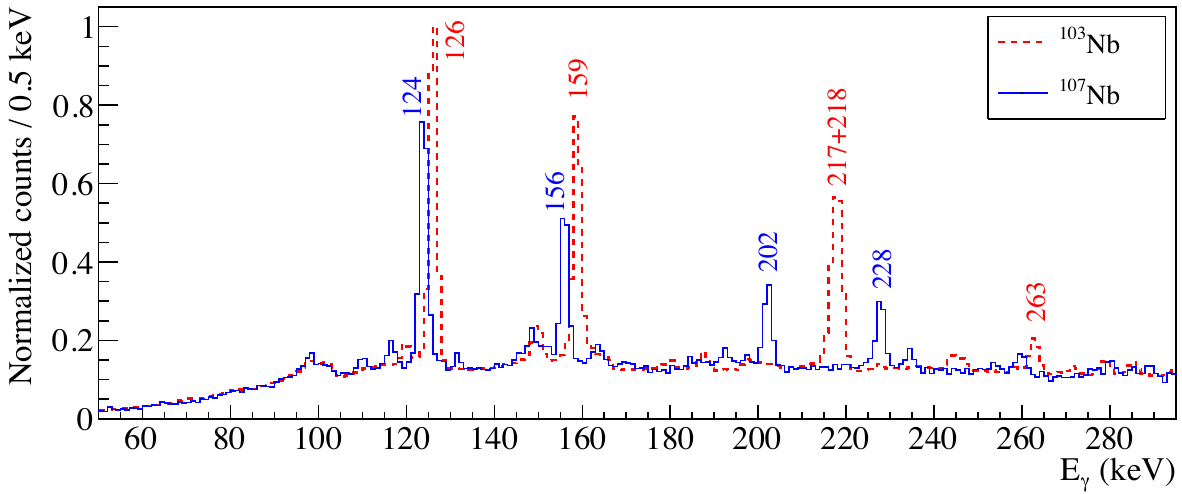}
\caption{Partial energy spectra of $^{103}$Nb (red dashed line) and $^{107}$Nb (solid blue line). The most intense transitions are labeled (see text).}
\label{fig:103vs107Nb}
\end{figure}

The high quality of isotopic identification and $\gamma$-ray energy resolution achieved in this work demonstrates that the transition energies published in~\cite{Hagen2017} (125, 158, 218 and 264~keV) are actually the result of a mixture of $^{107}$Nb and a contaminant arising from $^{103}$Nb. Figure~\ref{fig:103vs107Nb} displays the low-energy region of the $\gamma$ spectra of $^{103}$Nb (red dashed line) and $^{107}$Nb (solid blue line), with the most intense transitions labeled. Such an $(A-4)$ contamination is attributed to a $(Q-1)$ charge-state misidentification due to a lower energy resolution in~\cite{Hagen2017} (see~\cite{Lemasson2023} for details on the identification). This work therefore provides the first unambiguous level scheme for $^{107}$Nb.

In the present work, 13 new transitions associated with $^{107}$Nb were identified. The $\gamma$-ray spectrum of $^{107}$Nb is shown in Fig.~\ref{fig:singles} (e), with the new transitions labeled. The $\gamma$-ray transitions observed in this work, along with their relative intensities, are listed in Table~\ref{tab:107Nb}. The new level scheme is presented in Fig.~\ref{fig:105_107Nb_LS}.

\begin{table}
\caption{Level energies ($E_i$), $\gamma$-ray energies (E$_\gamma$), and relative intensities (I$_\gamma$) of transitions assigned to $^{107}$Nb in this work, along with the spin and parity of the initial ($J_i$) and final ($J_f$) states. Intensities are given relative to the strongest observed transition. New levels and transitions are in bold. Two transitions assigned to $^{107}$Nb that have not been placed on the level scheme are also reported.}
\begin{ruledtabular}
\begin{tabular}{lllcc}
$E_i$ (keV)             & E$_\gamma$ (keV)      & I$_\gamma$ $(\%)$     & $J_i^\pi$             &  $J_f^\pi$                \\
\hline  
0                       &                       &                       & (5/2+)                &                           \\
\textbf{124.0(2)}       & \textbf{124.0(2)}     & \textbf{100}          & $\mathbf{(7/2^+)}$   & $\mathbf{(5/2^+)}$        \\
\textbf{280.0(3)}       & \textbf{156.0(2)}     & \textbf{61(6)}        & $\mathbf{(9/2^+)}$   & $\mathbf{(7/2^+)}$        \\
\textbf{508.5(3)}       & \textbf{227.9(3)}     & \textbf{32(4)}        & $\mathbf{(11/2^+)}$  & $\mathbf{(9/2^+)}$        \\
                         & \textbf{385.1(4)}     & \textbf{9(3)}         & $\mathbf{(11/2^+)}$  & $\mathbf{(7/2^+)}$        \\
\textbf{710.3(3)}       & \textbf{202.3(3)}     & \textbf{37(4)}        & $\mathbf{(13/2^+)}$  & $\mathbf{(11/2^+)}$       \\
                         & \textbf{429.8(3)}     & \textbf{17(3)}        & $\mathbf{(13/2^+)}$  & $\mathbf{(9/2^+)}$        \\
\textbf{1038.1(3)}      & \textbf{327.3(3)}     & \textbf{15(3)}        & $\mathbf{(15/2^+)}$  & $\mathbf{(13/2^+)}$       \\
                         & \textbf{530.6(5)}     & \textbf{6(2)}         & $\mathbf{(15/2^+)}$  & $\mathbf{(11/2^+)}$       \\
\textbf{1273.1(3)}      & \textbf{563.0(5)}     & \textbf{13(3)}        & $\mathbf{(17/2^+)}$  & $\mathbf{(13/2^+)}$       \\
                         & \textbf{234.7(3)}     & \textbf{9(3)}         & $\mathbf{(17/2^+)}$  & $\mathbf{(15/2^+)}$       \\
\textbf{1726.0(4)}      & \textbf{687.8(6)}     & \textbf{9(3)}         & $\mathbf{(19/2^+)}$  & $\mathbf{(15/2^+)}$       \\
\multicolumn{5}{l}{\vspace{-8pt}} \\
\hline
\multicolumn{5}{l}{Unplaced transitions} \\
\hline
\multicolumn{5}{l}{\vspace{-8pt}} \\
-                        & \textbf{478.7(3)}     & \textbf{47(6)}        & -                     & -                         \\
-                        & \textbf{401.6(4)}     & \textbf{9(3)}         & -                     & -                         \\
\end{tabular}
\end{ruledtabular}
\label{tab:107Nb}
\end{table}

\begin{figure}
\includegraphics[width=\linewidth]{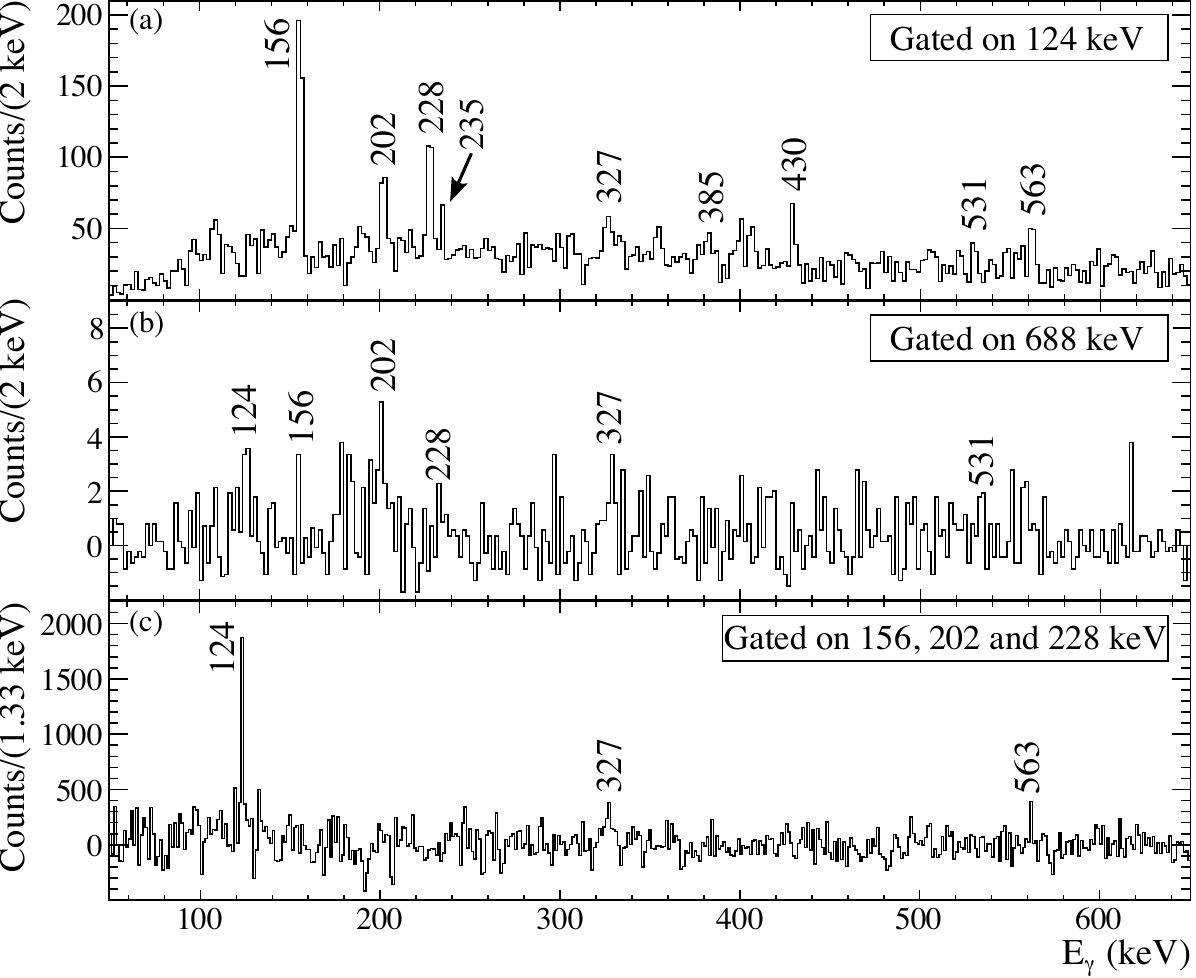}% 
\caption{AGATA $\gamma$-ray coincidence spectra gated on the 124~keV (a) and 688~keV (b) transitions of $^{107}$Nb. (c) Gammasphere triple-coincidence spectrum gated on the 156, 202, and 228~keV transitions in $^{107}$Nb.}
\label{fig:107Nb_gates}
\end{figure}

An example of coincidence spectra is shown in Fig.~\ref{fig:107Nb_gates} to confirm the newly proposed level scheme for $^{107}$Nb. Figure~\ref{fig:107Nb_gates} (a) presents the spectrum gated on the 124~keV transition, revealing coincidence with 156, 202, 228, 235, 327, 385, 430, 531, and 563~keV $\gamma$ rays. Figure~\ref{fig:107Nb_gates} (b) presents the spectrum gated on the 688~keV transition. Although the statistics are limited, the band transitions are observed in coincidence, confirming the placement of this transition at the top of the level scheme. This coincidence analysis is complemented by Gammasphere data, as shown on the triple-gated spectrum of Fig.~\ref{fig:107Nb_gates} (c). This combined analysis allowed the construction of a band composed of two interconnected cascades, similar to those observed in the ground-state bands of $^{101-105}$Nb. Therefore, the proposed spin assignments are based on the hypothesis of two E2 cascades linked through M1/E2 transitions, with a $(5/2^+)$ band head associated with the $\pi 5/2^+[422]$ Nilsson orbital, as previously suggested in~\cite{Hagen2017} (see Section~\ref{sec:interpretation} for more details).

Two transitions, isotopically assigned to $^{107}$Nb with energies of 402 and 479~keV, were not observed in coincidences with other transitions and were therefore not placed in the level scheme. Notably, the 479~keV transition is strongly populated, with a relative intensity of 47~\%. No equivalent transition of such high intensity, not connected to the main bands, has been observed in the lighter Nb isotopes.

\subsection{\texorpdfstring{$^{109}$}{109}Nb}

To our knowledge, the only publication reporting on the spectroscopy of $^{109}$Nb originates from an in-flight fission experiment using a $^{238}$U beam at the RIBF facility~\cite{Watanabe2011}. In this study, low-lying levels in $^{109}$Nb were populated via the $\beta$-decay of $^{109}$Zr, leading to the identification of a 313~keV isomer with a half-life of $T_{1/2} = 150(30)$~ns. Very recently, the spin of the $^{109}$Nb ground state was assigned to $5/2^{+}$ based on results from a dedicated $\beta$-decay experiment\cite{Bae2025}.

Statistically, $^{109}$Nb lies at the detection limit of the present work. The measured spectrum shown in Fig.~\ref{fig:singles} shows three transitions with energies of 116, 142, and 227~keV. The 116~keV transition was previously reported in~\cite{Watanabe2011}, whereas the other two are reported for the first time.

\begin{figure}
\includegraphics[width=\linewidth]{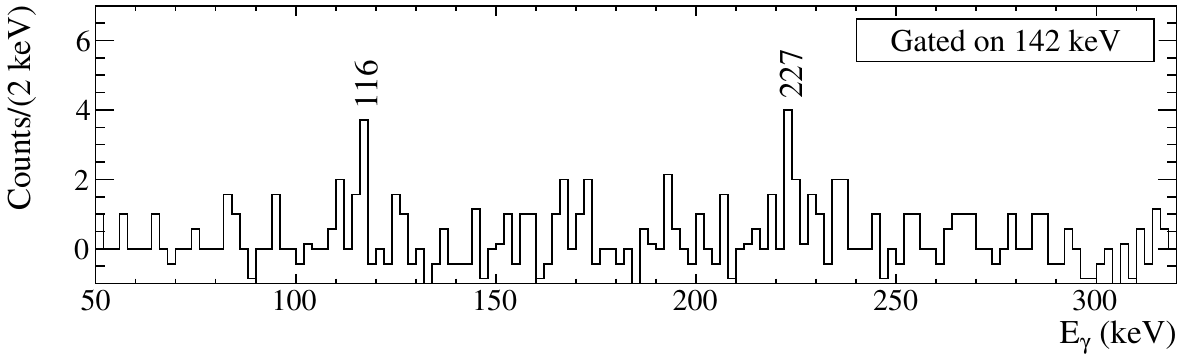}% 
\caption{Doppler-corrected $\gamma$-ray coincidence spectrum gated on the 142~keV transition of $^{109}$Nb measured in AGATA.}
\label{fig:109Nb_gates}
\end{figure}

A coincidence spectrum obtained with AGATA data, gated on the 142~keV transition, is shown in Fig.~\ref{fig:109Nb_gates}. It shows a clear coincidence with the 116 and 227~keV $\gamma$ rays. The $\gamma$-ray transitions observed in this work are listed in Table~\ref{tab:109Nb}, with new levels and transitions in bold. Owing to the limited statistics, reliable relative intensity values could not be determined with reasonable uncertainties. The updated level scheme is presented in Fig.~\ref{fig:105_107Nb_LS}, with new transitions and states highlighted in red. The spin assignment was determined based on systematic comparisons with lighter Nb isotopes (see Section~\ref{sec:interpretation}).

\begin{table}
\caption{Level energies ($E_i$) and $\gamma$-ray energies (E$_\gamma$) of transitions assigned to $^{109}$Nb in this work, along with the spin and parity of the initial ($J_i$) and final ($J_f$) states. New levels and transitions are highlighted in bold.}
\begin{ruledtabular}
\begin{tabular}{llcc}
$E_i$ (keV)             & E$_\gamma$ (keV)      & $J_i^\pi$             &  $J_f^\pi$                \\
\hline
0                       & -                     & $(5/2^+)$             & -                         \\
116.3(9)                & 116.3(9)              & $(7/2^+)$             & $(5/2^+)$                 \\
\textbf{258(2)}         & \textbf{141.7(9)}     & $\mathbf{(9/2^+)}$    & $\mathbf{(7/2^+)}$        \\
\textbf{485(2)}         & \textbf{227(1)}       & $\mathbf{(11/2^+)}$   & $\mathbf{(9/2^+)}$        \\
\end{tabular}
\end{ruledtabular}
\label{tab:109Nb}
\end{table}

\section{Systematic comparisons and interpretation of the level schemes}
\label{sec:interpretation}

\subsection{\texorpdfstring{$^{99}$}{99}Nb}

\subsubsection{Ground state positive-parity band}

\begin{figure}
\includegraphics[width=\linewidth]{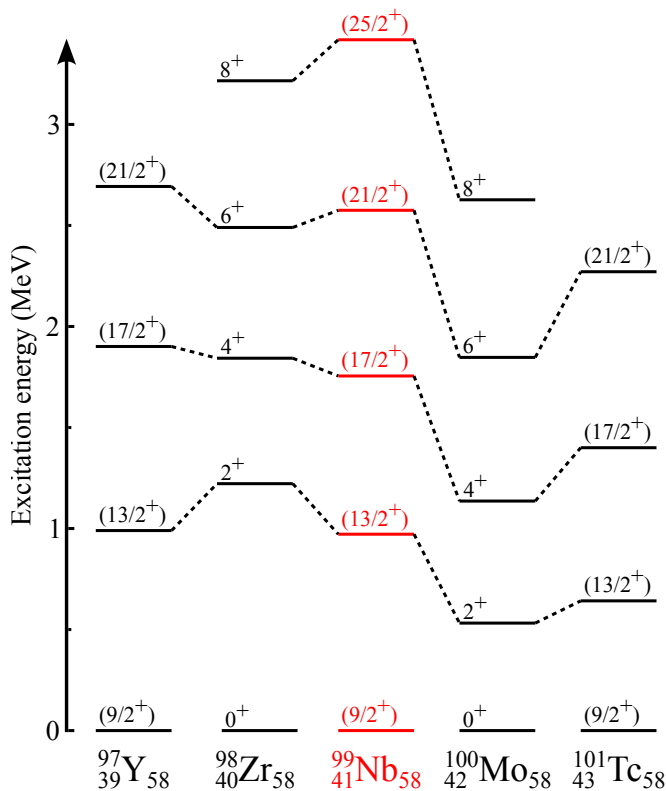}% 
\caption{Comparison of yrast states in the N=58 isotones from Y to Tc. For Y, energies are given relative to the $9/2^+$ state. Experimental data except that for $^{99}$Nb are taken from Ref.~\cite{ensdf}.}
\label{fig:99Nb_GS_syst}
\end{figure}

The newly reported ground-state band of $^{99}$Nb is compared to $N=58$ isotones, $^{97}$Y, $^{98}$Zr, $^{100}$Mo, and $^{101}$Tc in Fig.~\ref{fig:99Nb_GS_syst}. The energies and spin variations are given relative to the $9/2^+$ state for the odd-$Z$ nuclei and to the $0^+$ state for even-$Z$ nuclei. The energy levels of $^{99}$Nb appear to be more similar to those of $^{98}$Zr and $^{97}$Y than to those of $^{100}$Mo and $^{101}$Tc. This effect becomes even more pronounced in the revised level scheme of $^{99}$Nb proposed in this work, which incorporates new values for the $21/2^+$ and $25/2^+$ states. For these $N=58$ isotones, the level energy spacing remains relatively constant, which is characteristic of nuclei that can be approximated as spherical vibrators. This behavior is consistent with that of other $N=58$ nuclei in this region, which are only two neutrons away from the transition to highly deformed shapes.

The abrupt change in energy of the 392 and 361~keV $\gamma$ rays (see Fig.\ref{fig:99Nb_LS}), relative to the other transitions in the band, may indicate a strong structure modification at high spin (above $25/2^+$ state). As no systematic study confirms this observation and no calculations are available, no tentative spin assignments are proposed for these two levels at the top of the band.

A recent theoretical study by N. Gavrielov~\cite{Gavrielov2023} investigated the structure of $^{99}$Nb using the interacting boson-fermion model with configuration mixing (IBFM-CM). For the positive-parity states, the $13/2^+$ level is predicted at 944~keV (compared to 972~keV experimentally). However, the model does not predict a direct transition from this state to the ground state but instead suggests a low-energy transition to an $11/2^+$ state, which subsequently decays to a $9/2^+_2$ state — neither of which is observed experimentally.

\subsubsection{Excited negative-parity band}

The excited band of $^{99}$Nb, built on the $1/2^-$ isomeric state, is discussed in~\cite{Kumar2023}. In that study, this band was interpreted as a deformed structure built on the $5/2^-$ level at 630~keV and associated with the $\pi 5/2^-[303]$ Nilsson orbital, in analogy with the well-established bands in $^{101,103}$Nb and their odd-$A$ Y and Tc isotones. This interpretation led to a coherent picture of shape coexistence, whereby the same deformed band based on an isomeric state would persist from the transitional nucleus $^{99}$Nb ($N=58$) into the well-deformed region ($N\geq60$).

However, this revised interpretation of the level scheme questions this scenario. A strikingly similar band had already been identified in $^{101}$Tc~\cite{Savage1997,Hoellinger1999}, where it was shown, based on level energies, decay patterns, and particle-rotor calculations, to originate from the $\pi 1/2^-[301]$ Nilsson orbital, not from $\pi 5/2^-[303]$. In $^{101}$Tc, the band is built on a $1/2^-$ isomer, fed by a $3/2^-$ level, from which two parallel E2 cascades emerge, connected by M1/E2 transitions, as observed in $^{99}$Nb, including very similar transition energies.

\begin{figure}
\includegraphics[width=\linewidth]{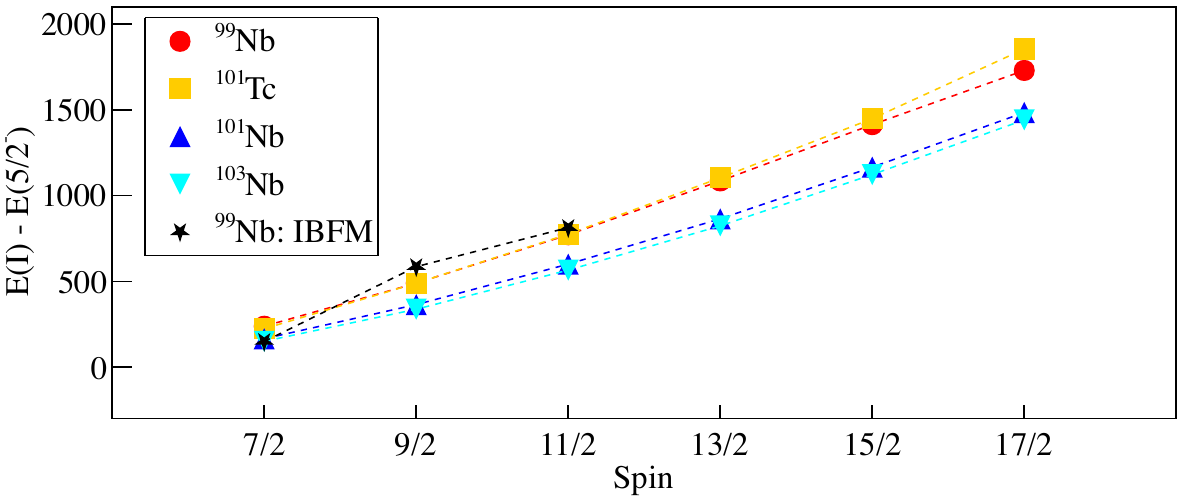}% 
\caption{Energies of the states in the negative-parity band observed in $^{99}$Nb with the $\pi 1/2^-[301]$ band of $^{101}$Tc, the $\pi 5/2^-[303]$ bands of $^{101}$Nb and $^{103}$Nb. IBFM-CM calculations~\cite{Gavrielov2023} are also shown. Energies are given relative to the $5/2^-$ state as a function of the spin. Experimental data except that for $^{99}$Nb are taken from Ref.~\cite{ensdf}.}
\label{fig:99Nb_DB_syst}
\end{figure}

Figure~\ref{fig:99Nb_DB_syst} compares this structure in $^{99}$Nb to the $\pi 1/2^-[301]$ band in $^{101}$Tc and the $\pi 5/2^-[303]$ bands in $^{101,103}$Nb. The similarities with $^{101}$Tc are evident and strongly support an assignment to the $\pi 1/2^-[301]$ configuration. This conclusion is further reinforced by recent IBFM-CM calculations~\cite{Gavrielov2023, Pfeil2023}, which predict two $K^\pi=1/2^-$ bands, one of which closely reproduces the experimental pattern observed in $^{99}$Nb, as shown in the figure. It is important to note that this band was unknown at the time of the IBFM-CM results~\cite{Pfeil2023,Gavrielov2023}.

The present results therefore indicate that, while the deformed bands in $^{101-103}$Nb are associated with the $\pi 5/2^-[303]$ configuration, the corresponding band in $^{99}$Nb originates from a different structure based on the $\pi 1/2^-[301]$ Nilsson orbital. This reveals a more complex situation than previously assumed, where bands with similar decay patterns and collective character arise from different underlying single-particle configurations on either side of the shape transition at $N=60$.

\subsection{\texorpdfstring{$^{101-109}$}{101-109}Nb: Energy systematics}

\begin{figure}
\includegraphics[width=\linewidth]{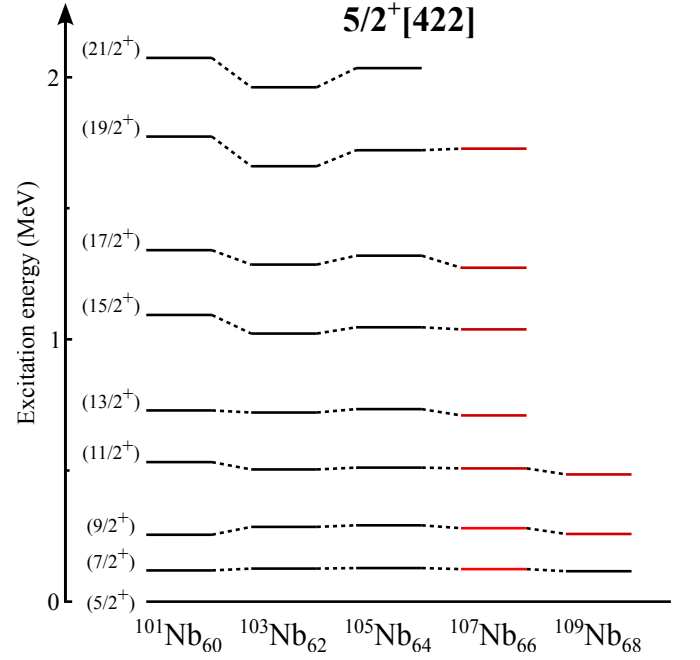}% 
\caption{Systematics of the level energies of the ground-state $\pi 5/2^+[422]$ band in $^{101-109}$Nb. New levels identified in this work are shown in red. Experimental data for $^{101-103}$Nb are taken from Ref.~\cite{ensdf}.}
\label{fig:Syst_422}
\end{figure}

The experimental data obtained in this work did not allow for the measurement of transition multipolarities or the spin assignment of the states. The spin values proposed in the level schemes are based solely on systematic comparisons with neighboring nuclei. Figure~\ref{fig:Syst_422} presents the energy systematics of levels assigned to the ground-state $\pi 5/2^+[422]$ band in $^{101-109}$Nb, with new levels from this work highlighted in red. Despite minor energy variations, the overall trend exhibits remarkable stability in energy levels from $N=60$ to $N=68$. Based on this observation, the new states reported in this work for the ground-state band of $^{107}$Nb and $^{109}$Nb are assigned to the $\pi 5/2^+[422]$ Nilsson orbital.

\begin{figure}
\includegraphics[width=\linewidth]{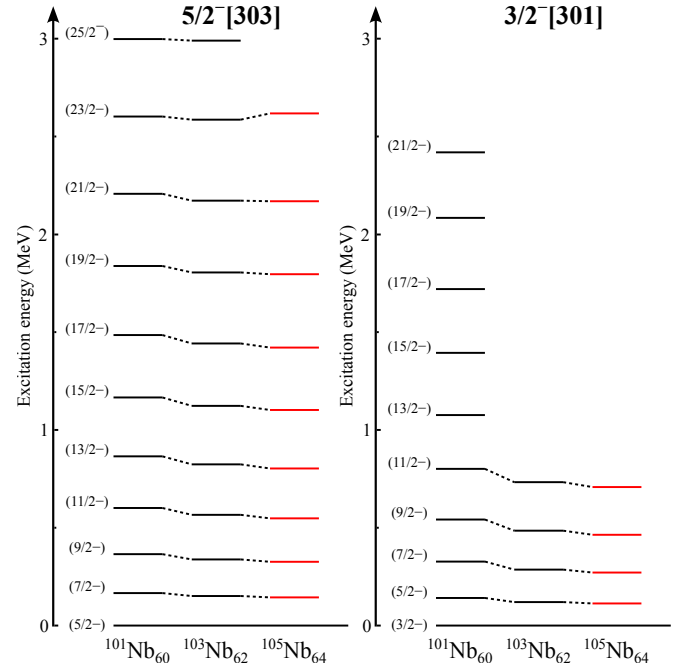}% 
\caption{Systematics of the level energies of negative-parity bands associated with the Nilsson orbitals $\pi 5/2^-[303]$ and $\pi 3/2^-[301]$ in $^{101-105}$Nb. New levels identified in this work are highlighted in red. Experimental data for $^{101-103}$Nb are taken from Ref.~\cite{ensdf}.}
\label{fig:Syst_303_301}
\end{figure}

Figure~\ref{fig:Syst_303_301} presents a similar comparison for the two negative-parity bands associated with the Nilsson orbitals $\pi 5/2^-[303]$ and $\pi 3/2^-[301]$ in $^{101-105}$Nb. The newly identified levels from this work in $^{105}$Nb are highlighted in red. Level energies are given relative to the $5/2^-$ and $3/2^-$ band-heads, respectively. Once again, a remarkably stable trend is observed, although a gradual decrease with neutron number suggests an increase in deformation. As with the ground-state band, this systematic trend is considered sufficiently strong to tentatively assign the spin and parity of the newly identified states in $^{105}$Nb.

This effect of stability in excitation energies has been discussed in~\cite{Kumar2021}. Starting from $N=60$, the onset of deformation leads to a stable configuration that persists with increasing neutron number, resulting in highly stable structures up to $N=70$ for Zr and up to $N=66$ for Mo.

Further experimental support for this analysis comes from the measured branching ratios. As illustrated in the level schemes in Fig.~\ref{fig:105_107Nb_LS}, the ground-state band is predominantly populated by $\Delta I=1$ transitions, whereas the negative-parity bands are dominated by $\Delta I=2$ transitions. This pattern is consistently observed in the present work for the odd-even $^{101}$Nb to $^{109}$Nb isotopes, as also reported in Refs.~\cite{Luo2005,Hagen2017}. Similar branching ratios have been measured in Y~\cite{Luo2005} and Tc~\cite{Luo2004} for these two bands.

\begin{figure}
\includegraphics[width=\linewidth]{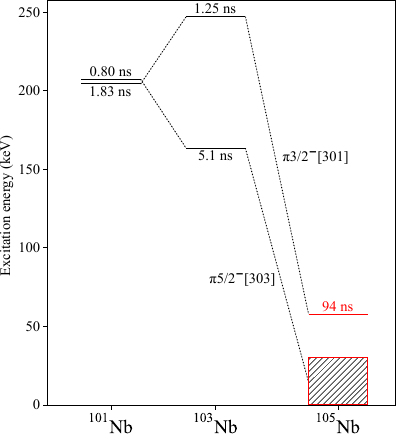}% 
\caption{Systematics of the energies and half-lives of the $\pi 5/2^-[303]$ and $\pi 3/2^-[301]$ band-heads. The shaded area indicates the range of the $\pi 5/2^-[303]$ predicted band-head energy from this work. Experimental data for $^{101-103}$Nb are taken from Ref.~\cite{ensdf}.}
\label{fig:Syst_303_301_BH}
\end{figure}

The present study provides the first insights into the negative-parity bands in Nb isotopes beyond $N=62$. Figure~\ref{fig:Syst_303_301_BH} illustrates the evolution of the $\pi 5/2^-[303]$ and $\pi 3/2^-[301]$ band-head energies and half-lives from $^{101}$Nb to $^{105}$Nb. A sudden decrease in energy, correlated with an increase in lifetime, is observed for the $3/2^-$ state in $^{105}$Nb, suggesting enhanced stability of this configuration. However, no transition to the ground state has been detected for the $\pi 5/2^-[303]$ configuration. The band-head energy level is therefore likely below the detection threshold of these experimental setups ($\sim30$~keV), represented by the red box in Fig.~\ref{fig:Syst_303_301_BH}. This energy decrease is likely associated with an increasing isomeric lifetime, as observed for the $3/2^-$ state.

In $^{103}$Nb, the transition from the $\pi 3/2^-[301]$ band-head to the ground state ($E=247.6(2)$~keV and $T_{1/2} = 1.25(4)$~ns) has been tentatively assigned an E1 multipolarity, corresponding to a reduced transition probability of $B(E1) = 1.62 \times 10^{-5}(6)$~W.u~\cite{NDS_A103}. Based on the new lifetime measurement of the corresponding state in $^{105}$Nb obtained in this work, and assuming the same E1 character, the deduced value for the $3/2^-\rightarrow 5/2^-$ transition is $B(E1) = 1.05\times10^{-5},^{+0.16}_{-0.12}$~W.u., including the effect of internal conversion. This value is consistent with a strongly hindered E1 decay and is of the same order of magnitude as that reported for $^{103}$Nb, supporting the interpretation of a similar underlying configuration and decay mode in both isotopes.

Conversely, the $B(E1)$ value observed in $^{103}$Nb can be used to estimate a range for the lower limit on the lifetime of the $\pi 5/2^-[303]$ band-head in $^{105}$Nb. The ground-state transition from the analogous state in $^{103}$Nb has an energy of $164.01(11)$~keV and a half-life of the emitting level of $T_{1/2} = 5.1(1)$~ns~\cite{NDS_A103}, corresponding to a $B(E1) = 1.33 \times 10^{-5}(3)$~W.u. Assuming a comparable $B(E1)$ value and a transition energy below 30~keV, this leads to an estimated minimum half-life on the order of 200~ns for the $\pi 5/2^-[303]$ state in $^{105}$Nb. This estimate is consistent with the interpretation of a long-lived isomeric state associated with this configuration.

\subsection{Nb: transition to triaxiality}

Studies of even-even nuclei with $N\ge60$ in this region have shown that $_{38}$Sr and $_{40}$Zr isotopes exhibit axially deformed shapes~\cite{Togashi2016,Ramos2019}. In contrast, $_{42}$Mo and $_{44}$Ru isotopes exhibit triaxial shapes~\cite{Abusara2017,Hagen2018,Kumar2021}. In this context, the transitional behavior of Nb isotopes regarding the onset of triaxiality has previously been studied and discussed~\cite{Luo2005}. Those findings suggest that odd-mass $_{41}$Nb isotopes exhibit a transitional nature, bridging the gap between the axially symmetric deformation observed in Y isotopes and the triaxial deformation in Tc and Rh isotopes. This transitional behavior in Nb isotopes is characterized by intermediate signature splitting values compared to their Y and Tc/Rh counterparts. This study can be further extended using the new experimental data presented in this work.

\subsubsection{Signature splitting and triaxiality}

In an odd-A nucleus, the signature quantum number is given by $\alpha _I$=1/2(-l)$^{I-1/2}$~\cite{Bengtsson1979}. In triaxially deformed nuclei, the $K$ is no longer a good quantum number and the band configurations become admixtures of wave functions with different $K$ components. The amount of admixture of the $\Omega = 1/2$ component governs the degree of signature splitting, which varies smoothly as a function of nuclear deformation and Fermi level. The sensitivity of signature splitting with triaxiality has been proven for Nb and other odd-Z isotopes in this region.

For strongly deformed coupled bands, the emergence of triaxiality induces irregular level energy spacings~\cite{Davydov1958,Zamfir1991}. This staggering in the energies can be observed in the energy systematics of the Nb ground-state band (see Fig.~\ref{fig:Syst_422}), where levels appear bunched together ($7/2^+ - 9/2^+$, $11/2^+ - 13/2^+$, etc.). Even more pronounced examples are found in Tc~\cite{Luo2006} and Rh isotopes~\cite{Luo2004,Hagen2018,Navin2017}, where triaxiality is known to be significant ($\gamma\sim30^\circ$). This staggering effect is typically analyzed using the signature splitting function $S(I)$, originally proposed by N.V. Zamfir and R.F. Casten~\cite{Zamfir1991,Casten1991}, which is defined as follows:

\begin{equation}
S(I) = \frac{R(E_I)}{R(E_I)_{Rotor}} - 1,
\end{equation}
with
\begin{equation}
R(E_I) = \frac{2\times[E_I-E_{I-1}]}{E_I-E_{I-2}}.
\end{equation}
and where $R(E_I)_{Rotor}$ represents $R(E_I)$ for a rigid axial rotor, given by $E(I)\propto I(I+1)$. This leads to the definition of the signature splitting S(I) used in this work:
\begin{equation}
S(I) = \frac{E_I - E_{I-1}}{E_I - E_{I-2}} \cdot \frac{I(I+1) - (I-2)(I-1)}{I(I+1) - (I-1)I} - 1.
\end{equation}

To aid in the interpretation of $S(I)$, the $R(E_I)_{Rotor}$ term normalizes the $S(I)$ plot around 0 for a perfect rigid rotor. If the rigid rotor assumption is not valid, the average trend will shift toward negative values due to imperfect normalization. In the presence of significant triaxiality, $E_I$ and $E_{I-1}$ (or $E_{I+1}$) become nearly degenerate, leading to $S(I)$ values approaching $\pm1$. For a purely axially symmetric shape, the band spacing remains uniform, yielding $S(I) = 0$.

\subsubsection{Ground-state band: \texorpdfstring{$\pi 5/2^+[422]$}{p5/2+[422]}}

The degree of triaxiality in $^{101-107}$Nb has been studied using triaxial particle-rotor calculations, initially up to $^{105}$Nb~\cite{Luo2005}, and more recently extended to $^{107}$Nb~\cite{Hagen2017}. It is worth noting that in this later study, the $^{107}$Nb level scheme was incorrect, as demonstrated in the present work (see Section~\ref{sec:107Nb}). Both studies indicate a nearly constant quadrupole deformation along the Nb isotopic chain ($\epsilon_2\sim0.37$) and a gradual decrease in triaxiality with increasing neutron number ($\gamma=19^\circ$ for $^{101}$Nb, $\gamma=15^\circ$ for $^{103}$Nb and $\gamma=13^\circ$ for $^{105}$Nb).

\begin{figure}
\includegraphics[width=\linewidth]{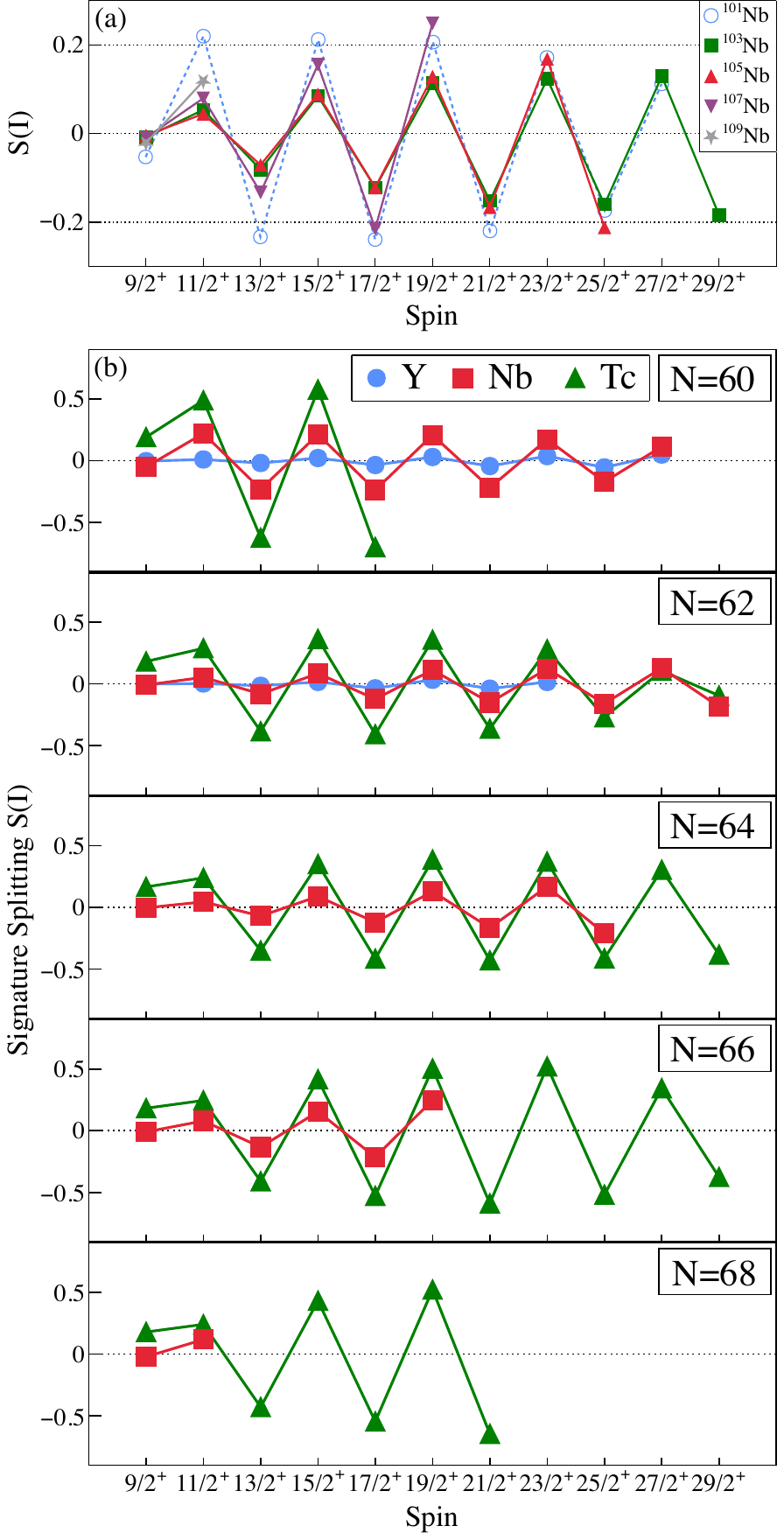}% 
\caption{Signature splitting, S(I), in the $\pi 5/2^+[422]$ band for the $N=60$ to $N=68$ isotones. (a) Systematics for the odd-even Nb isotopes. (b) Same as in (a) for ($Z=39$), Nb ($Z=41$) and Tc ($Z=43$) isotopes. Experimental data for $^{101-103}$Nb, Y and Tc are taken from Ref.~\cite{ensdf}.}
\label{fig:SI_422_Nb}
\end{figure}

Figure~\ref{fig:SI_422_Nb} (a) presents the experimental signature splitting of the $\pi 5/2^+[422]$ band in $^{101-109}$Nb isotopes. A distinct trend is observed between $^{101}$Nb and $^{103-109}$Nb. For $^{103-109}$Nb, a moderate degree of signature splitting is observed ($\sim\pm0.2$), with a slight increase in amplitude as the neutron number increases. The new results from this work, which incorporate $^{107-109}$Nb into this systematic study, suggest a more pronounced increase in triaxiality with neutron number. This observation contradicts the theoretical calculations discussed above~\cite{Luo2005,Hagen2017} which suggest a decrease in triaxiality with increasing neutron number. 

$^{101}$Nb deviates from the observed trend, exhibiting a nearly constant $S(I)$ amplitude even for the lowest levels of this band, unlike heavier Nb isotopes, where the amplitude increases with spin. This phenomenon has been studied in~\cite{Luo2005,Hagen2017}, where a value of $\gamma = 19^\circ$ was determined to reproduce the signature splitting at low spin. Applying this value to fit the low-spin region led to a divergence at high spin, where theoretical values were overestimated. However, within the broader systematics, it is observed that as spin increases, $S(I)$ values for $^{101}$Nb gradually decrease and become nearly identical to those of $^{103-105}$Nb. Indeed, in~\cite{Luo2005}, it was shown that to reproduce the high-spin states, a value of $\gamma = 14^\circ$ was required, which is consistent with the values calculated for $^{103-105}$Nb. In this context, $^{101}$Nb seems to exhibit a triaxiality decrease with spin, from $\gamma = 19^\circ$ at low spin to $\gamma = 14^\circ$ at high spin.

Having examined signature splitting along the Nb isotopic chain, it can now be compared as a function of $Z$ for different $N$ values. This comparison is illustrated in Fig.~\ref{fig:SI_422_Nb} (b) for Y ($Z=39$), Nb ($Z=41$) and Tc ($Z=43$) nuclei, covering $N=60$ to $N=68$. As discussed in~\cite{Luo2005}, this figure clearly illustrates the transition from axial shapes in Y nuclei to triaxial shapes in Tc, with an increase of approximately a factor of 10 in $S(I)$. Within this shape transition, the Nb isotopic chain serves as an intermediate regime toward triaxiality, exhibiting moderate $S(I)$ values. It should be noted that the signature splitting plots are all well centered around zero, indicating that the rigid rotor hypothesis used for normalization in the $S(I)$ formula is valid. This conclusion is consistent with the calculations presented in~\cite{Hagen2017}, which support the validity of the rotational model approximation. The new results from this work for $N=66$ and $N=68$ are consistent with this systematic trend.

\begin{figure}
\includegraphics[width=\linewidth]{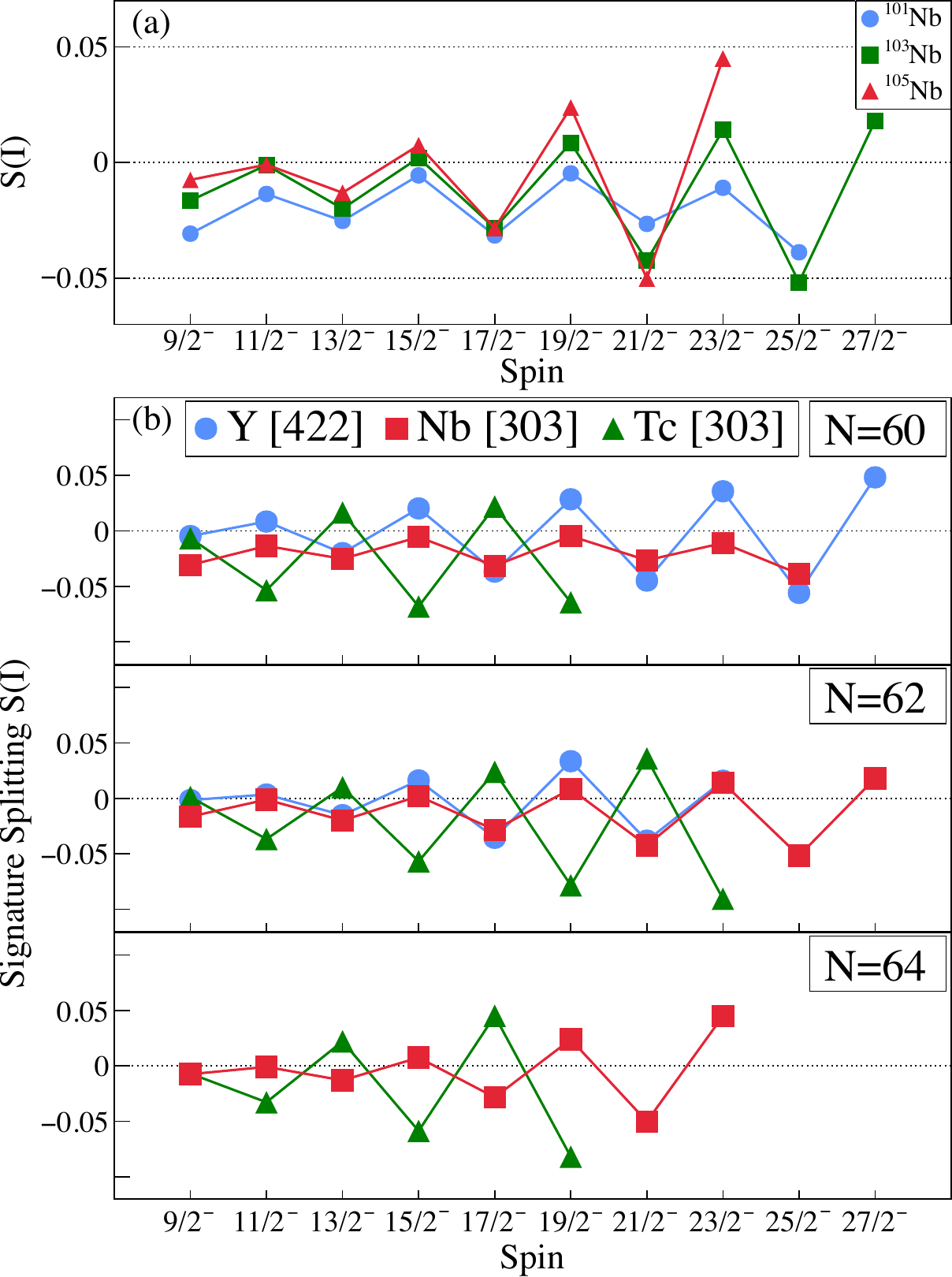}% 
\caption{Signature splitting, S(I), in the $\pi 5/2^-[303]$ band for the $N=60$ to $N=64$ isotones. It should be noted that the signature splitting values are a factor 10 smaller than in Fig.\ref{fig:SI_422_Nb}. (a) Odd-even Nb isotopes. (b) Nb ($Z=41$) and Tc ($Z=43$) isotopes, compared to the $\pi 5/2^+[422]$ band in Y ($Z=39$) isotopes. Experimental data for $^{101-103}$Nb, Y and Tc are taken from Ref.~\cite{ensdf}.}
\label{fig:SI_303_Nb}
\end{figure}

\begin{figure}
\includegraphics[width=\linewidth]{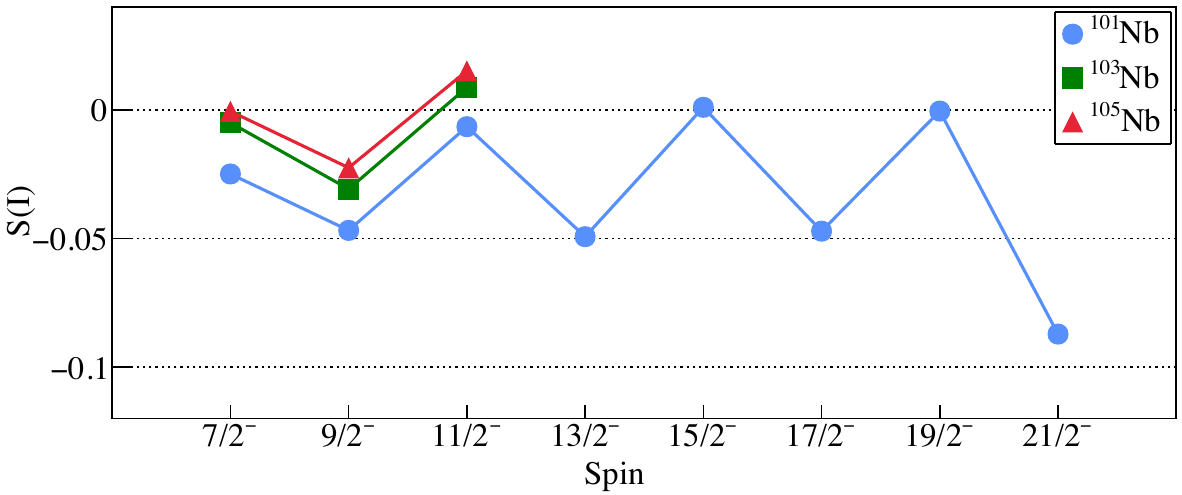}% 
\caption{Systematics of signature splitting, S(I), in the $\pi 3/2^-[301]$ band in Nb isotopes from $N=60$ to $N=64$. Experimental data for $^{101-103}$Nb are taken from Ref.~\cite{ensdf}.}
\label{fig:SI_301_Nb}
\end{figure}

\subsubsection{Negative-parity bands: \texorpdfstring{$\pi 5/2^-[303]$ and $\pi 3/2^-[301]$}{p5/2-[303] and p3/2-[301]}}

A similar comparison can then be made for the negative-parity bands. Figure~\ref{fig:SI_303_Nb} (a) presents the signature splitting of the $\pi 5/2^-[303]$ band for $^{101-105}$Nb isotopes. The first observation is that the $S(I)$ amplitude values are an order of magnitude lower than those of the previously discussed ground-state band and increase with both spin and neutron number. Furthermore, in contrast to what was observed for the ground-state band, the average value is no longer centered around zero. The assumption of a rigid rotor seems thus no longer valid.

As with the ground-state band, the $\pi 5/2^-[303]$ band can be analyzed as a function of $Z$ for different $N$ values. This comparison is illustrated in Fig.~\ref{fig:SI_303_Nb} (b) for Nb and Tc nuclei from $N=60$ to $N=64$. For comparison, the data for Y nuclei are also included in this figure, corresponding to the $\pi 5/2^+[422]$ ground-state band. The comparison between Nb and Tc reveals that although the values for Tc are slightly higher than for Nb, the $S(I)$ amplitude remains minimal, staying below 0.05. Moreover, Tc and Nb are out of phase, consistently persisting from $N=60$ to $N=64$. This implies that, even if there is almost no signature splitting, a very small energy difference (on the order of $\sim10-20$ keV) exists between the two coupled bands in Nb and Tc, inverting the staggering effect and persisting across all neutron numbers.

The amplitude of the measured signature splitting suggests that this band is axially deformed, in contrast to the triaxial ground-state band. Nevertheless, theoretical calculations reproducing these experimental observations are required for definitive conclusions on the deformation of this band. For comparison, the signature splitting of the ground-state band of Y isotopes, known to be purely axial, is also shown in Fig.~\ref{fig:SI_303_Nb} (b). The signature splitting values for the Y ground-state band closely match those of the $\pi 5/2^-[303]$ band in Nb. For $^{103}$Nb ($N=62$), the comparison is striking. Despite differences in the energy levels, the corresponding signature splitting is identical, suggesting that both exhibit an axially deformed structure.

Finally, a similar comparison can be made for the $\pi 3/2^-[301]$ band. Figure~\ref{fig:SI_301_Nb} illustrates this for $^{101-105}$Nb. The $S(I)$ amplitude falls within a similar range as that of the $\pi 5/2^-[303]$ band. Here again, the average value is not centered around zero, indicating that the rigid rotor approximation does not hold. Similar values are obtained for all Nb isotopes up to $11/2^-$ state. For $^{101}$Nb, the only isotope with a known structure above the $11/2^-$ state, a consistent oscillation is observed up to $11/2^-$ where a sudden drop in signature splitting suggests a structural modification at high spin.

\subsection{Shape coexistence}

To further support the coexistence of axial and triaxial shapes in Nb isotopes, a comparison is made with neighboring even-Z nuclei. Shape coexistence has been widely observed throughout the $Z\approx40$ region, where multiple low-lying structures of different shapes coexist and evolve with neutron number~\cite{Heyde2011,Garrett2022}. In this context, the odd-Z Nb isotopes offer an additional view on this phenomenon, where different proton configurations can couple to distinct core structures, resulting in coexisting rotational bands with different deformation characteristics.

\begin{figure*}
\includegraphics[width=\linewidth]{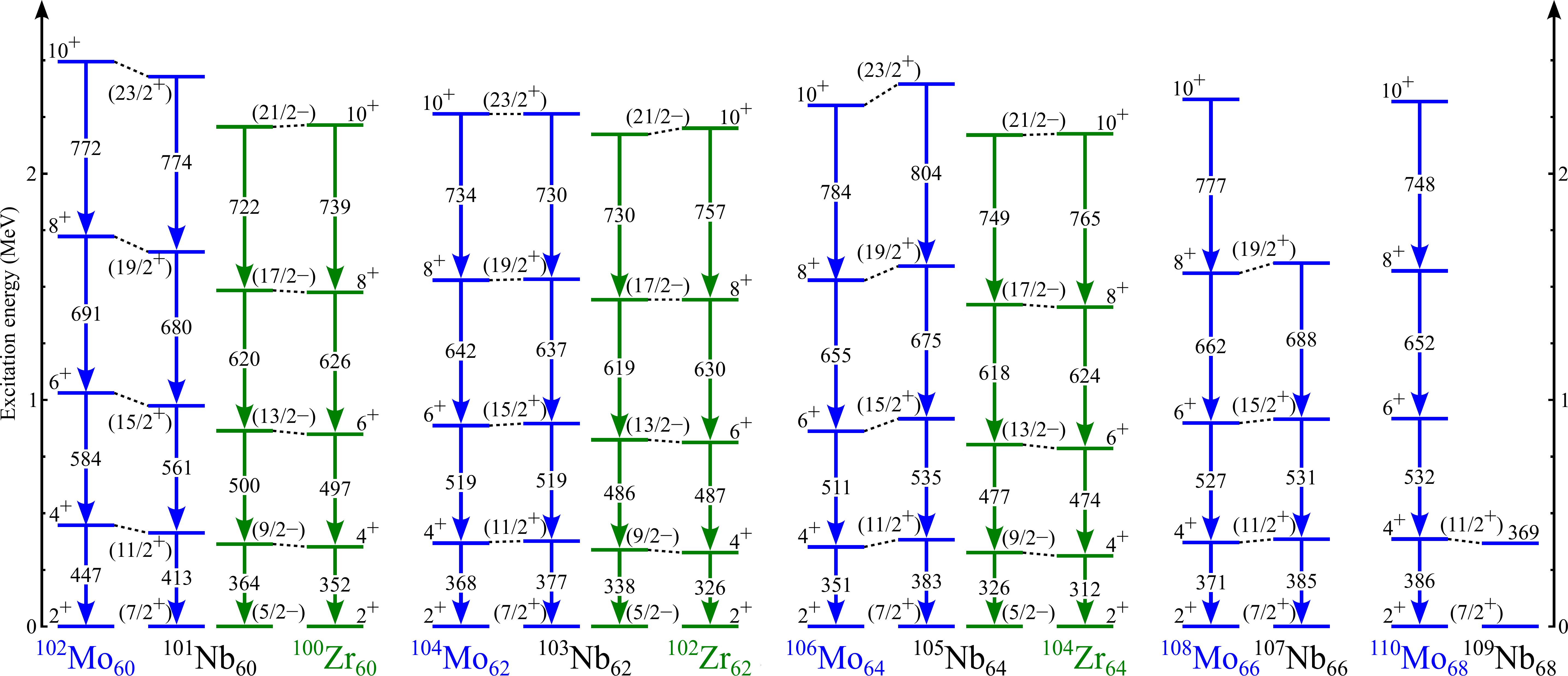}% 
\caption{Energies of the even-even Mo and Zr yrast band and bands of the neutron-rich odd-even Nb isotopes corresponding to the $\pi 5/2^+[422]$ and $\pi 5/2^-[303]$ Nilsson orbitals. To facilitate comparison, the lowest state is omitted. Dashed lines are included to guide the eye in the comparison. The 804~keV transition in $^{105}$Nb ($23/2^+ \rightarrow 19/2^+$) is taken from Ref.~\cite{Li2013}, as it was not observed in the present work. In $^{109}$Nb, the $11/2^+$ level has been identified, although the transition to $7/2^+$ has not been observed. Its energy is included here to ensure a consistent comparison. Experimental data for $^{101-103}$Nb, Mo and Zr are taken from Ref.~\cite{ensdf}.}
\label{fig:Syst_Mo_Zr}
\end{figure*}

Figure~\ref{fig:Syst_Mo_Zr} shows the levels and transitions in odd-even Nb nuclei and even-even Mo and Zr nuclei. To facilitate the comparison, first transition was omitted: an E2 transition in even-even nuclei, and an M1 (for the [422] band) or E1 (for the [303] band) transition in odd-even nuclei. Energy levels are thus relative to the $2^+$ state for Zr and Mo nuclei, and to the $5/2^+$ ($5/2^-$) state for the $\pi 5/2^+[422]$ ($\pi 5/2^-[303]$) bands in Nb. Only $\Delta I=2$ levels in Nb are considered for comparison with the E2 cascade in Zr and Mo.

This comparison highlights strong structural similarities between the ground-state bands of Nb and Mo, as well as between the negative-parity band of Nb and the ground-state band of Zr. The most striking case is $^{103}$Nb, where all energy differences remain below 30~keV. The experimental observations presented in this work suggest the coexistence of two distinct structures in Nb: a ground-state band with moderate triaxiality, structurally similar to higher-$Z$ nuclei such as Mo and Tc, and an excited band, based on a nanosecond-lifetime isomer associated with the $\pi 5/2^-[303]$ band, exhibiting axial deformation similar to lower-$Z$ nuclei like Zr and Y. The odd-mass Nb isotopes could therefore be interpreted as a proton hole coupled to a triaxially deformed Mo core in the ground-state configuration, coexisting with an excited configuration involving a proton particle coupled to the axially deformed Zr core.

\section{Conclusion}

In this work, new spectroscopic information for neutron-rich Nb isotopes was obtained through high-resolution $\gamma$-ray spectroscopy, combining two complementary experimental setups. The VAMOS++ spectrometer, coupled with AGATA, provided excellent isotopic selectivity and precise Doppler correction, enabling the identification $\gamma$-ray spectra up to the very neutron-rich $^{109}$Nb. These results were further enhanced by the high-statistics, high-fold coincidence data from Gammasphere, which confirmed key transitions and extended the level schemes to higher spins.

The level scheme of $^{99}$Nb has been revised, including a reinterpretation of the negative-parity band. For the positive-parity band, the level energy spacing remains nearly constant, underlining its vibrational character, consistent with that of the $N=58$ isotones. On the negative-parity side, the present study revises the previously proposed interpretation. The deformed band in $^{99}$Nb, initially associated with the $\pi 5/2^-[303]$ orbital as in heavier Nb isotopes, is shown to originate instead from the $\pi 1/2^-[301]$ configuration. This re-assignment, based on similarities with $^{101}$Tc and supported by IBFM-CM calculations, reveals a more complex picture: structurally similar bands in the $N = 58$ and $N \geq 60$ regions arise from different underlying single-particle configurations.

The systematic study of the band structure in the deformed region ($N\ge60$) has been extended up to $^{109}$Nb. A remarkable observation is the stability of the deformed bands built on the $\pi 5/2^+[422]$, $\pi 5/2^-[303]$, and $\pi 3/2^-[301]$ orbitals as a function of neutron number, from $N=60$ to $N=68$.

The analysis of signature splitting confirms the transitional nature of Nb isotopes between axially symmetric and triaxially deformed shapes. The $\pi 5/2^+[422]$ band exhibits signature splitting patterns indicative of increasing triaxiality with neutron number, in contrast with some previous theoretical predictions. Conversely, the $\pi 5/2^-[303]$ and $\pi 3/2^-[301]$ bands show significantly smaller splitting amplitudes, consistent with axial deformation.

The comparison of level energies between Nb and even-even Mo and Zr isotones reveals a new case of shape coexistence. The Nb ground-state band, with moderate triaxiality, closely resembles the rotational structures observed in Mo isotopes, while the negative-parity bands are structurally similar to the axial yrast band of Zr. This supports an interpretation of odd-A Nb isotopes as systems where a proton hole couples to a triaxial Mo-like core in the ground-state configuration, coexisting with an excited proton-particle configuration built on an axially deformed Zr-like core.

Altogether, this work refines our understanding of structural evolution in odd-Z nuclei in the $A\sim100$ mass region, providing a coherent framework that connects shape coexistence, triaxiality, and single-particle configurations. Future experimental efforts, including higher-statistics measurements and lifetime determinations beyond $N=64$, will extend these conclusions toward even more neutron-rich nuclei. In parallel, further theoretical developments are needed to provide reliable and predictive models for these complex systems, capable of describing odd-mass nuclei in regions of rapid shape transitions and configuration mixing.

\section*{Acknowledgments}

We would like to thank the E661 and E680 collaborations for their contributions to this work. The authors gratefully acknowledge the AGATA collaboration for the availability of the AGATA $\gamma$ tracking array at GANIL. We would like to thank the GANIL staff for their technical contributions. The authors are grateful to A.O.~Machiavelli for many fruitful discussions. This work at Shandong University is supported by the National Natural Science Foundation of China No.~12225504. The work at Vanderbilt University and Lawrence Berkeley National Laboratory are supported by the U.S. Department of Energy under Grant No.~DE-FG05-88ER40407 and Contract No.~DE-AC03-76SF00098. SB acknowledges the support received from CEFIPRA project No.~5604-4 and acknowledges the support from LIA France-India agreement.

\bibliography{biblio_20_authors}

\end{document}